# Modeling and control of permanent-magnet synchronous generators under open-switch converter faults

Christoph M. Hackl[‡,⋆], Urs Pecha[†] and Korbinian Schechner[‡]

**Abstract**

The mathematical modeling of open-switch faults in two-level machine-side converters and the fault-tolerant current control of isotropic permanent-magnet synchronous generators are discussed. The proposed converter model is generic for any open-switch fault and independent of the operation mode of the electrical machine. The proposed fault-tolerant current control system gives improved control performance and reduced torque ripple under open-switch faults by (i) modifying the anti-windup strategy, (ii) adapting the space-vector modulation scheme and (iii) by injecting additional reference currents. The theoretical derivations of model and control are validated by comparative simulation and measurement results.

**Index Terms**

Permanent-magnet synchronous generator, current control, fault-tolerance, fault-tolerant control, open-switch fault, anti-windup, field-oriented control, flat-top modulation, $d$-current injection, wind turbine systems

CONTENTS

[‡]C.M. Hackl and K. Schechner are with the research group “Control of renewable energy systems” (CRES) at the Munich School of Engineering (MSE), Technische Universität München (TUM), Germany.

[†]U. Pecha is with the Institute of Electrical Energy Conversion, University of Stuttgart, Germany.

[*]Authors are in alphabetical order and contributed equally to the paper. Corresponding author is C.M. Hackl (christoph.hackl@tum.de).

## Notation

$\mathbb{N}, \mathbb{R}, \mathbb{C}$: natural, real and complex numbers. $\mathbb{R}_{>\alpha} := (\alpha, \infty)$, $(\mathbb{R}_{\geq\alpha} := [\alpha, \infty))$: real numbers greater than (and equal to) $\alpha \in \mathbb{R}$. $\Re(s), \Im(s) \in \mathbb{R}$: real, imaginary part of $s \in \mathbb{C}$. $\boldsymbol{x} := (x_1, \ldots, x_n)^\top \in \mathbb{R}^n$: column vector, $n \in \mathbb{N}$ where '$^\top$' and ':=' mean 'transposed' and 'is defined as'. $\mathrm{diag}(a_1, \ldots, a_n) \in \mathbb{R}^{n \times n}$: diagonal matrix with entries $a_1, \ldots, a_n \in \mathbb{R}$, $n \in \mathbb{N}$. $\boldsymbol{I}_n := \mathrm{diag}(1, \ldots, 1) \in \mathbb{R}^{n \times n}$: identity matrix. $\boldsymbol{O}_{n \times p} \in \mathbb{R}^{n \times p}$: zero matrix, $n, p \in \mathbb{N}$. $\xi \overset{(\#1)}{\underset{(\#2)}{=}} \zeta$: equivalence of $\xi$ and $\zeta$ follows by invoking Eq. (#1) and Eq. (#2). $\boldsymbol{x} \in \mathbb{R}^n$ (in X)$^n$: physical quantity $\boldsymbol{x}$ where each of the $n$ elements has SI-unit X. $\mathrm{mod}\,(x, y)$: remainder of the division $x/y$, $x \in \mathbb{R}$, $y \in \mathbb{R}\backslash\{0\}$. atan2: $\mathbb{R}^2 \to [-\pi, \pi)$, $(x, y) \to \mathrm{atan2}(y, x)$ is the extension of the inverse tangent function to a whole circle. $\boldsymbol{T}_\mathrm{c} = \frac{2}{3}\begin{bmatrix} 1 & -\frac{1}{2} & -\frac{1}{2} \\ 0 & \frac{\sqrt{3}}{2} & -\frac{\sqrt{3}}{2} \end{bmatrix}$ and $\boldsymbol{T}_\mathrm{c}^{-1} = \frac{3}{2}\begin{bmatrix} \frac{2}{3} & 0 \\ -\frac{1}{3} & \frac{\sqrt{3}}{3} \\ -\frac{1}{3} & -\frac{\sqrt{3}}{3} \end{bmatrix}$: Clarke and inverse Clarke transformation matrix. $\boldsymbol{T}_\mathrm{p}(\phi) = \begin{bmatrix} \cos(\phi) & -\sin(\phi) \\ \sin(\phi) & \cos(\phi) \end{bmatrix} = \boldsymbol{T}_\mathrm{p}(-\phi)^{-1}$: Park transformation matrix. $\boldsymbol{J} = \begin{bmatrix} 0 & -1 \\ 1 & 0 \end{bmatrix}$: rotation matrix.

## I. Introduction

Open-switch faults in converters for electric drives have gained increasing attention in the last years. Especially, the detection of faults in the converter and the identification of the faulty switch have been the focus of research. Various detection methods have been presented [1], [2], [3], [4], [5], [6], [7], [8] and, hence, fault detection is *not* the topic of this paper.

In particular in off-shore wind turbine systems, it is desirable to continue their operation even in presence of open-switch faults in the machine-side converter. However, the faulty converter causes increased losses and oscillations in the torque which harm the mechanical components and the generator of the wind turbine [9]. To analyze the impact of open-switch faults, a model of the faulty converter has been proposed in [10], [11], [12]. The model determines the phase voltages of the electric machine connected to the faulty converter by using so called *pole voltages* of the converter. If there is an open-switch fault in one of the switching devices, a deviation in the pole voltage of the respective phase occurs which affects all three phase voltages. The pole voltages are, however, a rather unintuitive quantity compared to, for example, the *switching state* of the power electronic devices or the *phase voltages* of the machine. Moreover, the use of pole voltages in simulations makes the converter model unnecessarily complicated.

To ensure a safe and uninterrupted operation of the electrical drive, a fault-tolerant control strategy has to be implemented. In [13], [14], [15], [16], a modified space vector modulation (SVM) is proposed for two-level converters. These contributions adapt the switching patterns and replace those space vectors which cannot be applied due to the open-switch fault. However, these papers do neither consider optimal phase shift angles between applied voltage and current vector nor adapt their current controllers to the post-fault operation, although – as will be shown later – these measures additionally and significantly improve the overall control performance.

For three-level converters using NPC or T-type configurations the redundancy in the switching states can be used to compensate for the infeasible switching states and to generate the desired voltage output nevertheless, see [3], [5], [17], [18]. In addition, [18] proposes to inject an additional $d$-axis current to shift the phase angle between reference voltage and current to $0^\circ$ if an open-switch fault occurs in one of the outer switches. This helps to avoid infeasible switching states, and therefore reduces the current distortion in the faulty phase of the generator. However, the use of redundant switching vectors can not be used for two-level converters, since there is no sufficient redundancy in the available switching states. The impact of an open switch fault reduces the feasible voltage area in the voltage hexagon of a two-level converter significantly. Moreover, in [3], [5], [17], [18], the current control system and its impact on the control performance during faults are not discussed in detail.

[19] proposes to consider converters with open-switch faults as *three-switch three-phase rectifiers* (all upper or lower switches are assumed to be simply diodes). It investigates the possible avoidance of infeasible zero switching vectors in space vector modulation (SVM). To achieve a minimal current distortion, a phase shift of $180^\circ$ between current and voltage vector is proposed. This phase shift is achieved by injecting an appropriate $d$-current. Further investigations in this paper will show, that, for the considered PMSM, $180^\circ$ is not the *optimal* phase shift angle to guarantee a minimal current distortion. Moreover, in [19], the impact of the open-switch fault on the control performance of the current controller (such as windup effects) is neither addressed nor tackled, and a generic converter model covering open-switch faults for e.g. simulation purposes is not provided.

Another possibility to ensure a continued operation of the electric machine is the use of fault-tolerant converter configurations. Different topologies and methods to control faulty converters are described in [19], [20], [21], [22]. For example, in case of open-switch faults, a fourth inverter leg can be used, or the neutral point of the machine can be connected to the midpoint of the DC bus. Both solutions can compensate for the loss of the phase with the faulty switch. However, additional and costly hardware components or reconfigurations are required in contrast to standard converter configurations.

In this paper, a fault-tolerant current control system for PMSM-based wind turbine systems is proposed where the machine-side two-level converter exhibits open-switch faults. The proposed control system does *not* require any hardware modifications and is easy to implement, since the required extensions to the standard control system are straightforward and non-complex. Furthermore, a generic mathematical (phase) model of the faulty converter is proposed. The model relies on the *switching states* of the upper (or lower) switches and determines the phase voltages of the generator depending on the sign (direction) of the current in the faulty phase. Therefore, this model can be implemented quickly and efficiently. It is simple and easy to understand and allows very precise simulations which give almost identical results as the conducted experiments in the laboratory. Based on this precise but simple model, the impacts of the open-switch fault on the current control performance of a standard field-oriented control system are analyzed and a fault-tolerant control system is proposed. The proposed control system combines different modifications like (i) an improved anti-windup strategy, (ii) a modified SVM and (iii) an optimal $d$-axis current reference bringing the phase angle between voltage and current in the generator to an optimal value (which is neither $180^\circ$ nor $0^\circ$ for the considered machine). All modifications and their positive effects on the control performance under open-switch faults are discussed in detail. The effectiveness of the proposed modifications is finally illustrated and validated by comparative simulation and measurement results.

## II. Modelling

In this section, the models of the permanent-magnet synchronous machine (PMSM) and the machine-side inverter/converter with open-switch faults are introduced. The models are derived in the generic three-phase $(a, b, c)$-reference frame.

### A. Model of PMSM

The three-phase stator voltages of an isotropic PMSM are given by [23, Example 14.24]

$$\boldsymbol{u}_{\mathrm{s}}^{abc} = R_{\mathrm{s}}\boldsymbol{i}_{\mathrm{s}}^{abc} + \tfrac{\mathrm{d}}{\mathrm{d}t}\boldsymbol{\psi}_{\mathrm{s}}^{abc}(\boldsymbol{i}_{\mathrm{s}}^{abc}, \phi_{\mathrm{m}}) \tag{1}$$

with stator voltage (vector) $\boldsymbol{u}_{\mathrm{s}}^{abc} := (u_{\mathrm{s}}^{a}, u_{\mathrm{s}}^{b}, u_{\mathrm{s}}^{c})^\top$ (in $\mathrm{V})^3$, stator resistance $R_{\mathrm{s}}$ (in $\Omega$), stator current (vector) $\boldsymbol{i}_{\mathrm{s}}^{abc} := (i_{\mathrm{s}}^{a}, i_{\mathrm{s}}^{b}, i_{\mathrm{s}}^{c})^\top$ (in $\mathrm{A})^3$ and stator flux linkage (vector) $\boldsymbol{\psi}_{\mathrm{s}}^{abc} := (\psi_{\mathrm{s}}^{a}, \psi_{\mathrm{s}}^{b}, \psi_{\mathrm{s}}^{c})^\top$ (in $\mathrm{V\,s})^3$. Note that the stator phase currents sum up to zero (i.e. $i_{\mathrm{s}}^{a} + i_{\mathrm{s}}^{b} + i_{\mathrm{s}}^{c} = 0$) due to the star connection of the stator windings. The stator flux linkage

$$\boldsymbol{\psi}_{\mathrm{s}}^{abc}(\boldsymbol{i}_{\mathrm{s}}^{abc}, \phi_{\mathrm{m}}) = \underbrace{\begin{bmatrix} L_{\mathrm{s},m} + L_{\mathrm{s},\sigma} & -\frac{L_{\mathrm{s},m}}{2} & -\frac{L_{\mathrm{s},m}}{2} \\ -\frac{L_{\mathrm{s},m}}{2} & L_{\mathrm{s},m} + L_{\mathrm{s},\sigma} & -\frac{L_{\mathrm{s},m}}{2} \\ -\frac{L_{\mathrm{s},m}}{2} & -\frac{L_{\mathrm{s},m}}{2} & L_{\mathrm{s},m} + L_{\mathrm{s},\sigma} \end{bmatrix}}_{=:\boldsymbol{L}_{\mathrm{s}}^{abc}} \boldsymbol{i}_{\mathrm{s}}^{abc} + \underbrace{\hat{\psi}_{\mathrm{pm}} \begin{pmatrix} \cos\left(n_{\mathrm{p}}\left(\phi_{\mathrm{m}} + \phi_{\mathrm{pm}}\right)\right) \\ \cos\left(n_{\mathrm{p}}\left(\phi_{\mathrm{m}} + \phi_{\mathrm{pm}}\right) - \frac{2}{3}\pi\right) \\ \cos\left(n_{\mathrm{p}}\left(\phi_{\mathrm{m}} + \phi_{\mathrm{pm}}\right) - \frac{4}{3}\pi\right) \end{pmatrix}}_{=:\boldsymbol{\psi}_{\mathrm{pm}}^{abc}(\phi_{\mathrm{m}})}, \tag{2}$$

depends on the inductance matrix $\boldsymbol{L}_{\mathrm{s}}^{abc}$ (in $\frac{\mathrm{V\,s}}{\mathrm{A}})^3$ with stator main inductance $L_{\mathrm{s},m}$ & stator leakage inductance $L_{\mathrm{s},\sigma}$ (both in $\frac{\mathrm{V\,s}}{\mathrm{A}}$), stator currents $\boldsymbol{i}_{\mathrm{s}}^{abc}$ and permanent-magnet (PM) flux linkage vector $\boldsymbol{\psi}_{\mathrm{pm}}^{abc}$ with PM-flux linkage amplitude $\hat{\psi}_{\mathrm{pm}}$ (in $\mathrm{V\,s}$), number $n_{\mathrm{p}}$ of pole pairs, machine (mechanical) angle $\phi_{\mathrm{m}} := \int \omega_{\mathrm{m}}(\tau)\mathrm{d}\tau$ (in rad) and (initial) angle $\phi_{\mathrm{pm}}$ (in rad) of the permanent magnet. Inserting (2) into (1) yields the current dynamics combined with the mechanical dynamics in the $(a, b, c)$-reference frame (see Fig. 1) as follows

$$\left.\begin{aligned} \tfrac{\mathrm{d}}{\mathrm{d}t}\boldsymbol{i}_{\mathrm{s}}^{abc}(t) &= \left(\boldsymbol{L}_{\mathrm{s}}^{abc}\right)^{-1}\left(\boldsymbol{u}_{\mathrm{s}}^{abc}(t) - R_{\mathrm{s}}\boldsymbol{i}_{\mathrm{s}}^{abc}(t) + \underbrace{n_{\mathrm{p}}\omega_{\mathrm{m}}(t)\hat{\psi}_{\mathrm{pm}}\begin{pmatrix} \sin\left(n_{\mathrm{p}}\left(\phi_{\mathrm{m}}(t) + \phi_{\mathrm{pm}}\right)\right) \\ \sin\left(n_{\mathrm{p}}\left(\phi_{\mathrm{m}}(t) + \phi_{\mathrm{pm}}\right) - \frac{2}{3}\pi\right) \\ \sin\left(n_{\mathrm{p}}\left(\phi_{\mathrm{m}}(t) + \phi_{\mathrm{pm}}\right) - \frac{4}{3}\pi\right) \end{pmatrix}}_{=:-\boldsymbol{e}_{\mathrm{s}}^{abc} = -(e_{\mathrm{s}}^{a}, e_{\mathrm{s}}^{b}, e_{\mathrm{s}}^{c})^\top}\right) \\ \tfrac{\mathrm{d}}{\mathrm{d}t}\omega_{\mathrm{m}}(t) &= \tfrac{1}{\Theta}\left(m_{\mathrm{m}}\big(\boldsymbol{i}_{\mathrm{s}}^{abc}(t), \phi_{\mathrm{m}}(t)\big) + m_{\mathrm{t}}(t)\right) \\ \tfrac{\mathrm{d}}{\mathrm{d}t}\phi_{\mathrm{m}}(t) &= \omega_{\mathrm{m}}(t), \end{aligned}\right\} \tag{3}$$

with initial currents $\boldsymbol{i}_{\mathrm{s}}^{abc}(0) = \boldsymbol{i}_{\mathrm{s},0}^{abc}$ (in $\mathrm{A})^3$, initial angular velocity $\omega_{\mathrm{m}}(0) = \omega_{\mathrm{m},0}$ (in $\frac{\mathrm{rad}}{\mathrm{s}}$), initial machine angle $\phi_{\mathrm{m}}(0) = \phi_{\mathrm{m},0}$ (in rad), induced back-emf voltage vector $\boldsymbol{e}_{\mathrm{s}}^{abc} = (e_{\mathrm{s}}^{a}, e_{\mathrm{s}}^{b}, e_{\mathrm{s}}^{c})^\top$ (in $\mathrm{V})^3$, total inertia $\Theta$ (in $\mathrm{kg\,m}^2$) of the drive train, machine torque $m_{\mathrm{m}}$ and turbine torque $m_{\mathrm{t}}$ (both in $\mathrm{N\,m}$). The machine torque $m_{\mathrm{m}}$ can be computed as follows [23, Example 14.24]

$$m_{\mathrm{m}}(\boldsymbol{i}_{\mathrm{s}}^{abc}, \phi_{\mathrm{m}}) = \frac{n_{\mathrm{p}}}{3}\left(\boldsymbol{i}_{\mathrm{s}}^{abc}\right)^\top \begin{bmatrix} 1 & 1-\sqrt{3} & 1+\sqrt{3} \\ 1+\sqrt{3} & 1 & 1-\sqrt{3} \\ 1-\sqrt{3} & 1+\sqrt{3} & 1 \end{bmatrix} \boldsymbol{\psi}_{\mathrm{s}}^{abc}(\boldsymbol{i}_{\mathrm{s}}^{abc}, \phi_{\mathrm{m}}) = -n_{\mathrm{p}}\hat{\psi}_{\mathrm{pm}} \begin{pmatrix} i_{\mathrm{s}}^{a}\sin\left(n_{\mathrm{p}}\left(\phi_{\mathrm{m}}(t) + \phi_{\mathrm{pm}}\right)\right) \\ i_{\mathrm{s}}^{b}\sin\left(n_{\mathrm{p}}\left(\phi_{\mathrm{m}}(t) + \phi_{\mathrm{pm}}\right) - \frac{2}{3}\pi\right) \\ i_{\mathrm{s}}^{c}\sin\left(n_{\mathrm{p}}\left(\phi_{\mathrm{m}}(t) + \phi_{\mathrm{pm}}\right) - \frac{4}{3}\pi\right) \end{pmatrix}. \tag{4}$$

**Remark II.1** (Field orientation)**.** *Aligning the synchronously rotating $(d, q)$-reference frame with the PM-flux linkage, i.e. applying the Clarke and Park transformation $\boldsymbol{x}^k = \boldsymbol{T}_{\mathrm{p}}(\phi_k)^{-1}\boldsymbol{T}_{\mathrm{c}}\boldsymbol{x}^{abc}$ with $\phi_k = \int \omega_k + \phi_{\mathrm{pm}}$ and $\omega_k = n_{\mathrm{p}}\omega_{\mathrm{m}}$ to machine*

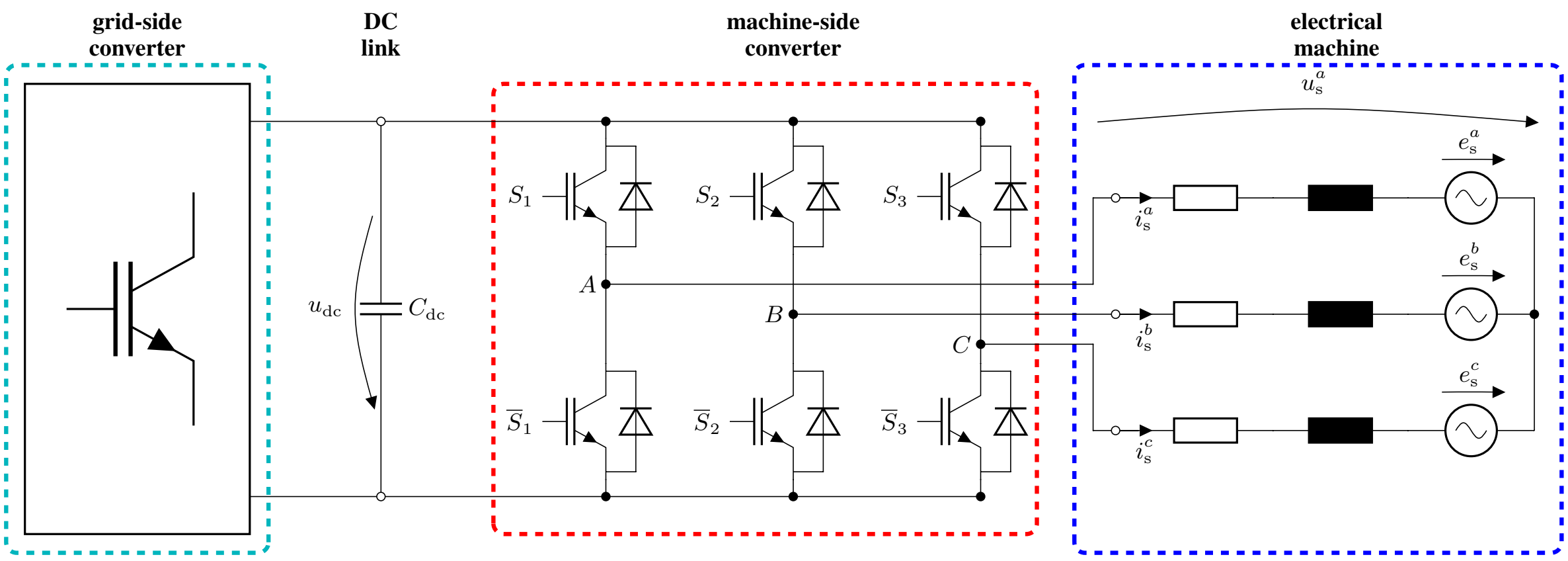

Fig. 1: *Back-to-back converter with permanent-magnet synchronous machine.*

*dynamics* (3) *and machine torque* (4) *yields the machine dynamics*

$$\left.\begin{array}{rcll} \frac{\mathrm{d}}{\mathrm{d}t}\boldsymbol{i}_{\rm s}^k(t) &=& \frac{1}{L_{\rm s}}\left(\boldsymbol{u}_{\rm s}^k(t) - R_{\rm s}\boldsymbol{i}_{\rm s}^k(t) - \omega_k(t)L_{\rm s}\boldsymbol{J}\boldsymbol{i}_{\rm s}^k(t) - \omega_k(t)\hat{\psi}_{\rm pm}\begin{pmatrix}0\\1\end{pmatrix}\right), & \boldsymbol{i}_{\rm s}^k(0) = \boldsymbol{T}_{\rm p}^{-1}(\phi_k(0))\boldsymbol{T}_{\rm c}\boldsymbol{i}_{\rm s,0}^{abc} \\ \frac{\mathrm{d}}{\mathrm{d}t}\omega_{\rm m}(t) &=& \frac{1}{\Theta}\left(m_{\rm m}\big(\boldsymbol{i}_{\rm s}^k(t)\big) + m_{\rm t}(t)\right), & \omega_{\rm m}(0) = \omega_{\rm m,0} \\ \frac{\mathrm{d}}{\mathrm{d}t}\phi_{\rm m}(t) &=& \omega_{\rm m}(t) & \phi_{\rm m}(0) = \phi_{\rm m,0} \end{array}\right\} \quad (5)$$

*in the* PM-flux linkage orientation *(or, simply,* field orientation*; see e.g. [23, Chapter 14] or [24, Sect. 3.2.2]) with stator voltages* $\boldsymbol{u}_{\rm s}^k := (u_{\rm s}^d,\ u_{\rm s}^q)^\top$ *(in* $\mathrm{V})^2$*, stator currents* $\boldsymbol{i}_{\rm s}^k := (i_{\rm s}^d,\ i_{\rm s}^q)$ *(in* $\mathrm{A})^2$*, stator inductance* $L_{\rm s} := \frac{3}{2}L_{{\rm s},m} + L_{{\rm s},\sigma}$ *(in* $\frac{\mathrm{V\,s}}{\mathrm{A}}$*) and flux linkage* $\boldsymbol{\psi}_{\rm s}^k := (\psi_{\rm s}^d,\ \psi_{\rm s}^q)^\top$ *(in* $\mathrm{V\,s})^2$*, and the machine torque*

$$m_{\rm m}(\boldsymbol{i}_{\rm s}^k) = \tfrac{3}{2}n_{\rm p}(\boldsymbol{i}_{\rm s}^k)^\top\boldsymbol{J}\boldsymbol{\psi}_{\rm s}^k = \tfrac{3}{2}n_{\rm p}\hat{\psi}_{\rm pm}i_{\rm s}^q. \quad (6)$$

### *B. Model of converter*

*1) Model of converter without faults:* Figure 1 shows a back-to-back converter with permanent-magnet synchronous machine. The output voltage of the machine-side converter is the stator voltage, given by [25]

$$\boldsymbol{u}_{\rm s}^{abc}(u_{\rm dc}, \boldsymbol{s}_{\rm s}^{abc}) = \frac{u_{\rm dc}}{3}\begin{bmatrix} 2 & -1 & -1 \\ -1 & 2 & -1 \\ -1 & -1 & 2 \end{bmatrix}\boldsymbol{s}_{\rm s}^{abc}, \quad (7)$$

and depends in the fault free case for a star-connected, symmetrical electrical machine only on the actual switching vector $\boldsymbol{s}_{\rm s}^{abc} = (s_{\rm s}^a,\ s_{\rm s}^b,\ s_{\rm s}^c)^\top$ and the actual dc-link voltage $u_{\rm dc}$ (in V). A "1" in the switching vector $\boldsymbol{s}_{\rm s}^{abc}$ means that the upper switch is closed. A "0" represents a closed lower switch. For example, a switching vector $\boldsymbol{s}_{\rm s}^{abc} = (1,\ 1,\ 0)^\top$ yields the closed switches $S_1$, $S_2$, and $\overline{S}_3$ ($S_3$ is open). Applying the Clarke transformation to (7) allows to transform $\boldsymbol{u}_{\rm s}^{abc}$ to the two-dimensional stator fixed $(\alpha, \beta)$-reference frame as follows $\boldsymbol{u}_{\rm s}^{\rm s} := \left(u_{\rm s}^\alpha,\ u_{\rm s}^\beta\right)^\top = \boldsymbol{T}_{\rm c}\boldsymbol{u}_{\rm s}^{abc}$.

*2) Model of converter with one open-switch fault:* At first, the model is derived for an open-switch fault in $S_1$. Afterwards, the generalization of the faulty converter model is presented. Without loss of generality, the open-switch fault is assumed to appear in switch $S_1$. Hence, switch $S_1$ is always open independent of the switching vector $\boldsymbol{s}_{\rm s}^{abc}$ of the converter. With this fault present, the voltage $\boldsymbol{u}_{\rm s}^{abc}$ does not solely depend on the dc-link voltage $u_{\rm dc}$ and the switching vector $\boldsymbol{s}_{\rm s}^{abc}$, but also on the sign (direction) of the current $i_{\rm s}^a$ in phase $a$. Figure 2(a) illustrates the different connection possibilities of the windings of the electrical machine taking the sign of the current into account and whether the free-wheeling diode of $S_1$ is conducting or not (compare also with Fig. 1). The resulting (shifted) voltage hexagon for this case is shown in Fig. 2(b). The following observations can be made:

- For $i_{\rm s}^a = 0$, the voltage vectors (see blue symbols) are shifted by $\boldsymbol{\Delta u}^s_{i_{\rm s}^a=0} = -\frac{1}{3}u_{\rm dc}$ in negative $a$-direction (see e.g. $\boldsymbol{u}^s_{110,i_{\rm s}^a=0}$ or $\boldsymbol{u}^s_{101,i_{\rm s}^a=0}$ in Fig. 2(b)). Normal operation is *not* feasible.
- For $i_{\rm s}^a > 0$, the voltage vectors (see magenta symbols) with a "1" for $S_1$ are shifted by $\boldsymbol{\Delta u}^s_{i_{\rm s}^a>0} = -\frac{2}{3}u_{\rm dc}$ in negative $a$-direction (see e.g. $\boldsymbol{u}^s_{101,i_{\rm s}^a>0}$ or $\boldsymbol{u}^s_{101,i_{\rm s}^a>0}$ in Fig. 2(b)). Normal operation is *not* feasible.
- For $i_{\rm s}^a < 0$, the voltage vectors (see green symbols) are *not* shifted (see e.g. $\boldsymbol{u}^s_{101,i_{\rm s}^a<0}$ or $\boldsymbol{u}^s_{110,i_{\rm s}^a<0}$ in Fig. 2(b)). Normal operation is feasible.

In Fig. 3, the feasible voltage areas in the voltage hexagon and the feasible voltage vectors $\boldsymbol{u}_{\rm s}^{\rm s}$ for open-switch faults in $S_1$ are shown depending on the direction of current $i_{\rm s}^a$. For $i_{\rm s}^a < 0$, the full voltage hexagon can be used. For $i_{\rm s}^a = 0$ or $i_{\rm s}^a > 0$, the feasible areas in the voltage hexagon become smaller. Most critical case occurs for $i_{\rm s}^a > 0$, where only the sectors III and IV are feasible. Combining the observations above, the converter model (7) must be extended for an open-switch fault in $S_1$

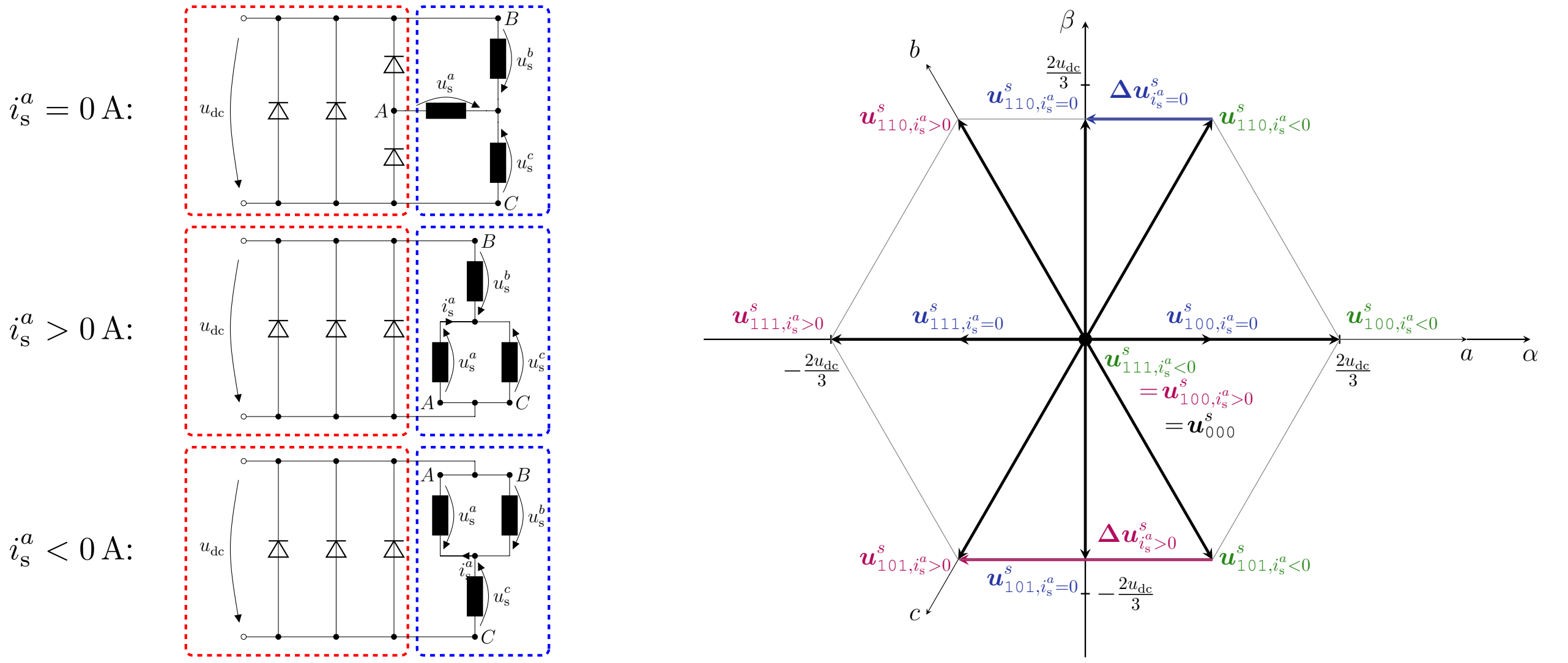

(a) Electrical equivalent circuit for $\boldsymbol{s}_s^{abc} = (1,\ 1,\ 0)^\top$ .

(b) Voltage hexagon with *shifted* voltage vectors for all possible $\boldsymbol{s}_s^{abc}$.

Fig. 2: *Illustration of impact of open-switch fault in $S_1$ on (a) electrical equivalent circuit of machine-side converter and PMSM (for simplicity, the windings are shown as inductances) and on (b) voltage hexagon with shifted voltage vectors $\boldsymbol{u}_s^s$ for $i_s^a = 0$ and $i_s^a > 0$.*

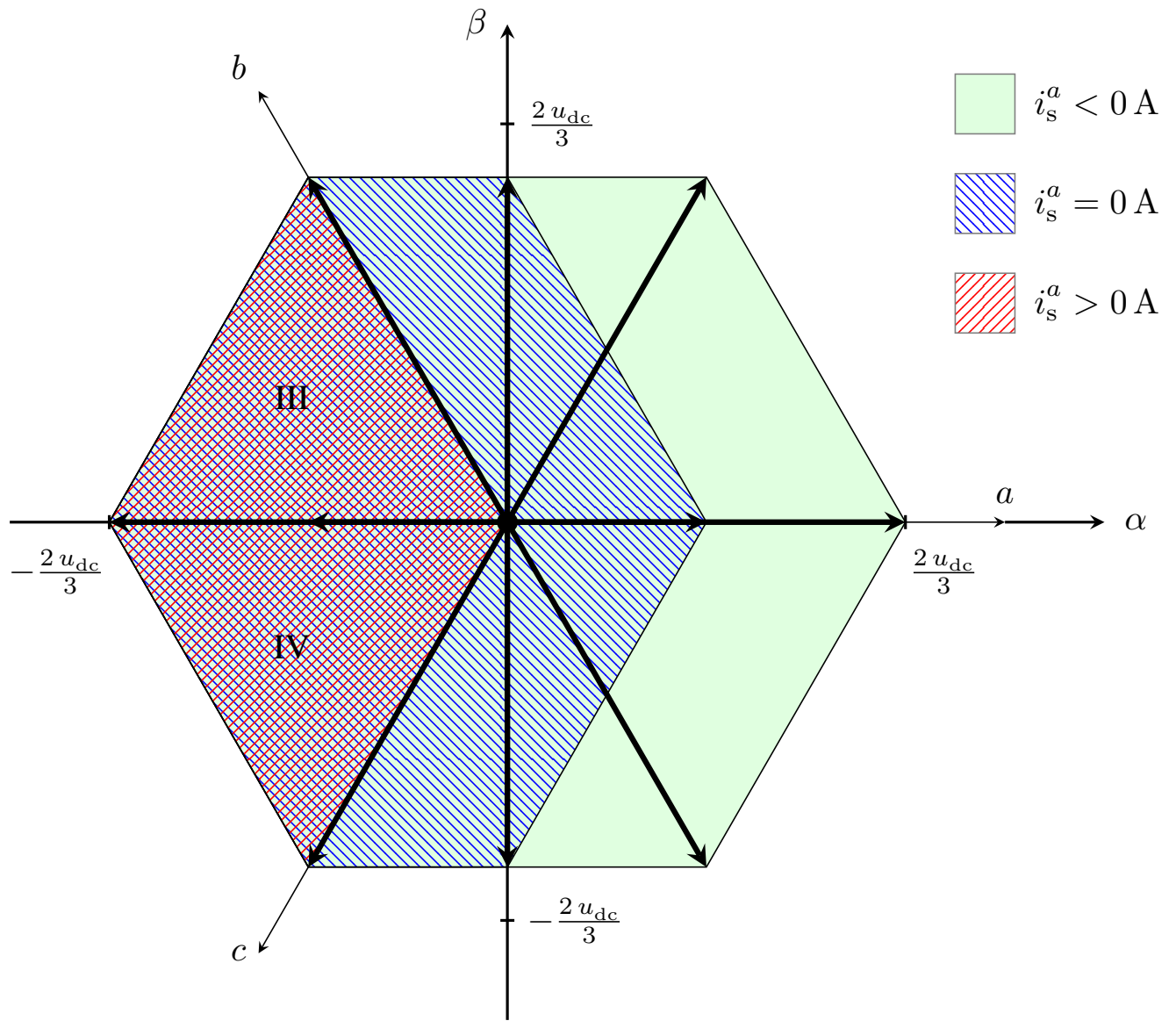

Fig. 3: *Voltage hexagon with feasible sectors and voltage vectors under an open-switch fault in $S_1$ depending on the current $i_s^a$.*

as follows

$$\boldsymbol{u}_s^{abc} = \frac{u_{dc}}{3} \underbrace{\left( \begin{bmatrix} 2 & -1 & -1 \\ -1 & 2 & -1 \\ -1 & -1 & 2 \end{bmatrix} + \begin{cases} \boldsymbol{O}_{3\times3}, & \text{if } i_s^a < 0\,\text{A} \\ \begin{bmatrix} -1 & 0 & 0 \\ \frac{1}{2} & 0 & 0 \\ \frac{1}{2} & 0 & 0 \end{bmatrix}, & \text{if } i_s^a = 0\,\text{A} \\ \begin{bmatrix} -2 & 0 & 0 \\ 1 & 0 & 0 \\ 1 & 0 & 0 \end{bmatrix}, & \text{if } i_s^a > 0\,\text{A} \end{cases} \right)}_{=:\boldsymbol{S}_{S_1}(i_s^a)} \boldsymbol{s}_s^{abc}. \tag{8}$$

The switching matrix $\boldsymbol{S}_{S_1}(i_s^a) \in \mathbb{R}^{3\times3}$ exclusively models the converter for an open-switch fault in $S_1$ and changes with the direction of the phase current $i_s^a$. Generalizing the observations above to an arbitrary open-switch fault in one of the six switches $S_1$, $S_2$, $S_3$, $\overline{S}_1$, $\overline{S}_2$ and $\overline{S}_3$, and introducing the negated switching vector

$$\overline{\boldsymbol{s}}_s^{abc} := \boldsymbol{1}_3 - \boldsymbol{s}_s^{abc} \text{ where } \boldsymbol{1}_3 := (1,\ 1,\ 1)^\top, \tag{9}$$

TABLE I: *Complete model of a faulty converter with switching matrices $\boldsymbol{S}_y$ and $\boldsymbol{S}_z$ for faulty switches $S_1, S_2, S_3$ and $\overline{S}_1, \overline{S}_2, \overline{S}_3$, resp.*

| | Converter output phase voltage $\boldsymbol{u}_{\mathrm{s}}^{abc} = \frac{u_{\mathrm{dc}}}{3}\boldsymbol{S}_y(i_{\mathrm{s}}^x)\,\boldsymbol{s}_{\mathrm{s}}^{abc}$ for faults in the *upper* switches | | |
|---|---|---|---|
| | $y = S_1, x = a$ | $y = S_2, x = b$ | $y = S_3, x = c$ |
| $\boldsymbol{S}_y\left(i_{\mathrm{s}}^x < 0\,\mathrm{A}\right)$ | $\begin{bmatrix} 2 & -1 & -1 \\ -1 & 2 & -1 \\ -1 & -1 & 2 \end{bmatrix}$ | $\begin{bmatrix} 2 & -1 & -1 \\ -1 & 2 & -1 \\ -1 & -1 & 2 \end{bmatrix}$ | $\begin{bmatrix} 2 & -1 & -1 \\ -1 & 2 & -1 \\ -1 & -1 & 2 \end{bmatrix}$ |
| $\boldsymbol{S}_y\left(i_{\mathrm{s}}^x = 0\,\mathrm{A}\right)$ | $\begin{bmatrix} 1 & -1 & -1 \\ -\frac{1}{2} & 2 & -1 \\ -\frac{1}{2} & -1 & 2 \end{bmatrix}$ | $\begin{bmatrix} 2 & -\frac{1}{2} & -1 \\ -1 & 1 & -1 \\ -1 & -\frac{1}{2} & 2 \end{bmatrix}$ | $\begin{bmatrix} 2 & -1 & -\frac{1}{2} \\ -1 & 2 & -\frac{1}{2} \\ -1 & -1 & 1 \end{bmatrix}$ |
| $\boldsymbol{S}_y\left(i_{\mathrm{s}}^x > 0\,\mathrm{A}\right)$ | $\begin{bmatrix} 0 & -1 & -1 \\ 0 & 2 & -1 \\ 0 & -1 & 2 \end{bmatrix}$ | $\begin{bmatrix} 2 & 0 & -1 \\ -1 & 0 & -1 \\ -1 & 0 & 2 \end{bmatrix}$ | $\begin{bmatrix} 2 & -1 & 0 \\ -1 & 2 & 0 \\ -1 & -1 & 0 \end{bmatrix}$ |
| | Converter output phase voltage $\boldsymbol{u}_{\mathrm{s}}^{abc} = \frac{u_{\mathrm{dc}}}{3}\boldsymbol{S}_z(i_{\mathrm{s}}^x)\,\overline{\boldsymbol{s}}_{\mathrm{s}}^{abc}$ for faults in the *lower* switches | | |
| | $z = \overline{S}_1, x = a$ | $z = \overline{S}_2, x = b$ | $z = \overline{S}_3, x = c$ |
| $\boldsymbol{S}_z\left(i_{\mathrm{s}}^x < 0\,\mathrm{A}\right)$ | $\begin{bmatrix} 0 & 1 & 1 \\ 0 & -2 & 1 \\ 0 & 1 & -2 \end{bmatrix}$ | $\begin{bmatrix} -2 & 0 & 1 \\ 1 & 0 & 1 \\ 1 & 0 & -2 \end{bmatrix}$ | $\begin{bmatrix} -2 & 1 & 0 \\ 1 & -2 & 0 \\ 1 & 1 & 0 \end{bmatrix}$ |
| $\boldsymbol{S}_z\left(i_{\mathrm{s}}^x = 0\,\mathrm{A}\right)$ | $\begin{bmatrix} -1 & 1 & 1 \\ \frac{1}{2} & -2 & 1 \\ \frac{1}{2} & 1 & -2 \end{bmatrix}$ | $\begin{bmatrix} -2 & \frac{1}{2} & 1 \\ 1 & -1 & 1 \\ 1 & \frac{1}{2} & -2 \end{bmatrix}$ | $\begin{bmatrix} -2 & 1 & \frac{1}{2} \\ 1 & -2 & \frac{1}{2} \\ 1 & 1 & -1 \end{bmatrix}$ |
| $\boldsymbol{S}_z\left(i_{\mathrm{s}}^x > 0\,\mathrm{A}\right)$ | $\begin{bmatrix} -2 & 1 & 1 \\ 1 & -2 & 1 \\ 1 & 1 & -2 \end{bmatrix}$ | $\begin{bmatrix} -2 & 1 & 1 \\ 1 & -2 & 1 \\ 1 & 1 & -2 \end{bmatrix}$ | $\begin{bmatrix} -2 & 1 & 1 \\ 1 & -2 & 1 \\ 1 & 1 & -2 \end{bmatrix}$ |

leads to different switching matrices $\boldsymbol{S}_y(i_{\mathrm{s}}^x)$ and $\boldsymbol{S}_z(i_{\mathrm{s}}^x)$ with $x \in \{a, b, c\}$, $y \in \{S_1, S_2, S_3\}$ and $z \in \{\overline{S}_1, \overline{S}_2, \overline{S}_3\}$, respectively. Finally, the switching matrices for all possible faults in the six switches $S_1$, $S_2$, $S_3$, $\overline{S}_1$, $\overline{S}_2$ and $\overline{S}_3$ and the respective phase current directions are collected in Table I (details are omitted due to space limitations).

## III. Current Control System

PI controllers and field-oriented control are a common choice for the current control system in electrical drives (see [26, Sect. 14.6] or [24, Sect. 3.2.2]). In most cases, the PI controllers are equipped with an anti-windup strategy and cross-coupling feedforward compensation terms to compensate for the cross-coupling between $d-$ and $q-$currents. In this section, at first, standard field-oriented control is briefly revisited. Afterwards, the crucial modifications to improve the control performance under open-switch faults are proposed. The impact of an open-switch fault on the standard control system and the improvements achieved by the proposed fault-tolerant control system are illustrated and analyzed in simulations (see Sect. III-A5 and Sect. III-B, resp.). Finally, the simulation results are validated by comparative measurement results in Sect. IV.

### A. Standard control system (field-oriented control)

In Fig. 4, the block diagram of the standard control system consisting of 1) PI controllers with anti wind-up, 2) cross-coupling feedforward compensation, 3) reference voltage saturation and 4) modulation is depicted. In the following subsections, each block will be explained briefly.

*1) PI controllers with anti-windup:* The PI controllers weight and integrate the current control error

$$\boldsymbol{e}_{\boldsymbol{i}_s}^k(t) := \begin{pmatrix} e_{\boldsymbol{i}_s}^d(t) \\ e_{\boldsymbol{i}_s}^q(t) \end{pmatrix} := \begin{pmatrix} i_{\mathrm{s,ref}}^d(t) - i_{\mathrm{s}}^d(t) \\ i_{\mathrm{s,ref}}^q(t) - i_{\mathrm{s}}^q(t) \end{pmatrix}, \tag{10}$$

defined as the difference between reference currents (coming e.g. from outer control loops) and the actual currents. Output $\boldsymbol{u}_{\mathrm{s,PI}}^k = (u_{\mathrm{s,PI}}^d, u_{\mathrm{s,PI}}^q)^\top$ and dynamics of integrator $\boldsymbol{\xi}_{\mathrm{i}}^k = (\xi_{\mathrm{i}}^d, \xi_{\mathrm{i}}^q)^\top$ of the PI controllers with anti-windup decision function $f_{\mathrm{aw}}(\cdot)$ are as follows

$$\left.\begin{aligned} \boldsymbol{u}_{\mathrm{s,PI}}^k(t) &= \begin{bmatrix} k_{\mathrm{p}}^d & 0 \\ 0 & k_{\mathrm{p}}^q \end{bmatrix} \boldsymbol{e}_{\boldsymbol{i}_s}^k(t) + \begin{bmatrix} k_{\mathrm{i}}^d & 0 \\ 0 & k_{\mathrm{i}}^q \end{bmatrix} \boldsymbol{\xi}_{\mathrm{i}}^k(t) \\ \tfrac{\mathrm{d}}{\mathrm{d}t}\boldsymbol{\xi}_{\mathrm{i}}^k(t) &= f_{\mathrm{aw}}\big(\hat{u}_{\mathrm{s,ref}}(t)\big) \cdot \boldsymbol{e}_{\boldsymbol{i}_s}^k(t), \qquad \boldsymbol{\xi}_{\mathrm{i}}^k(0) = \boldsymbol{0}_2. \end{aligned}\right\} \tag{11}$$

with proportional and integral controller gains $k_{\mathrm{p}}^d$ & $k_{\mathrm{p}}^q$ and $k_{\mathrm{i}}^d$ & $k_{\mathrm{i}}^q$, respectively. For example, a model-based tuning is according to the *magnitude optimum* (see e.g. [25]) which leads the following controller gains $k_{\mathrm{p}}^d = k_{\mathrm{p}}^q = \frac{L_{\mathrm{s}} f_{\mathrm{sw}}}{3}$ and $k_{\mathrm{i}}^d = k_{\mathrm{i}}^q = \frac{R_{\mathrm{s}} f_{\mathrm{sw}}}{3}$ where $f_{\mathrm{sw}}$ (in Hz) is switching frequency of the converter/inverter. Any other reasonable tuning rule might also be applicable. The anti-windup decision function

$$f_{\mathrm{aw}}(\hat{u}_{\mathrm{s,ref}}) := \begin{cases} 1 & \text{, if } \hat{u}_{\mathrm{s,ref}} \le \hat{u}_{\max}(u_{\mathrm{dc}}, \theta') \\ 0 & \text{, else} \end{cases} \tag{12}$$

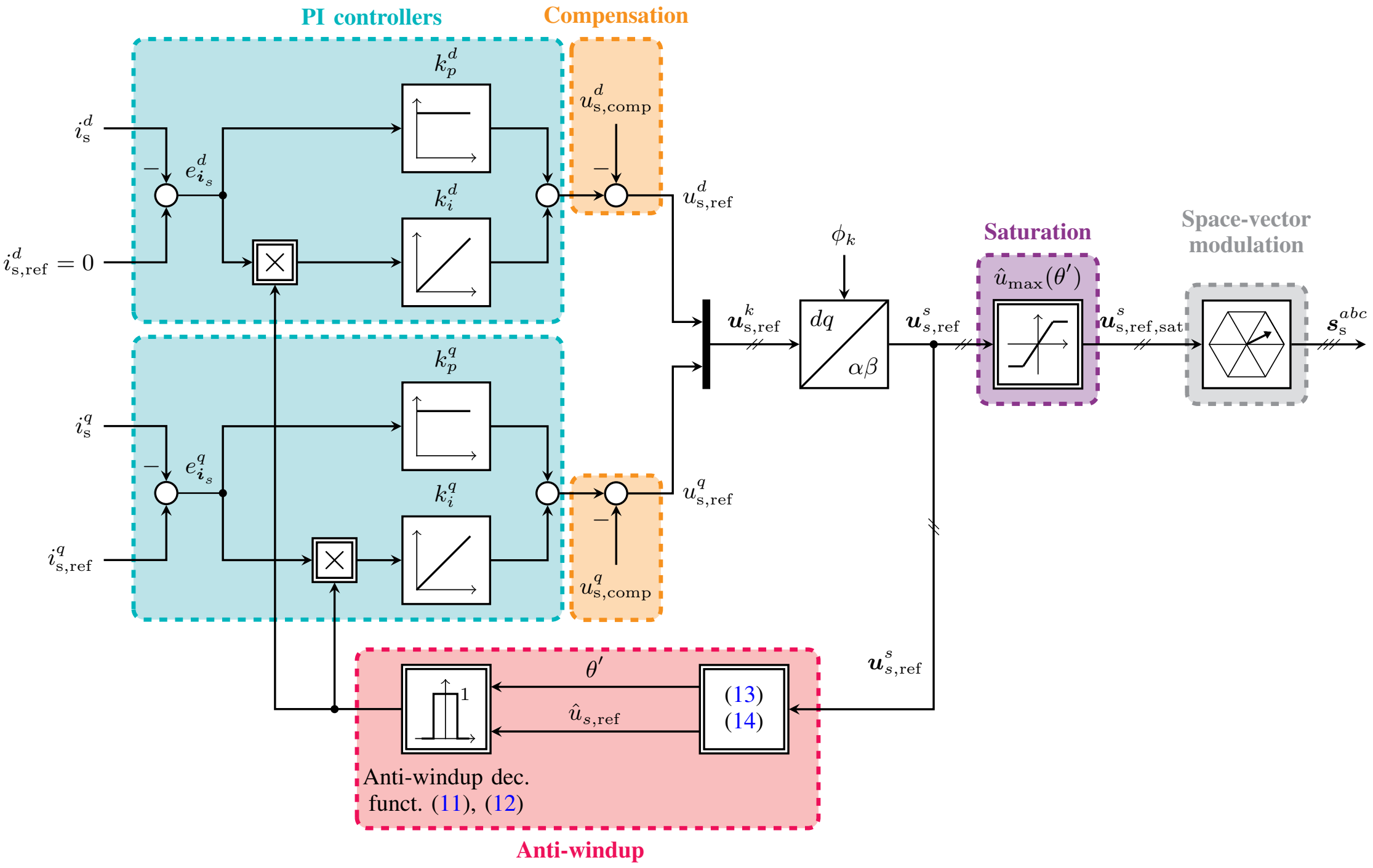

Fig. 4: *Standard control system (field-oriented control): PI controllers with anti-windup, cross-coupling feedforward compensation and reference voltage saturation.*

enables or disables integration of the integral control action (i.e. conditional integration, for more details see [23, Sect. 10.4.1]), if the applied reference voltage vector amplitude $\hat{u}_{\mathrm{s,ref}}$ (in V), defined by

$$\hat{u}_{\mathrm{s,ref}} := \sqrt{\left(u^{\alpha}_{\mathrm{s,ref}}\right)^2 + \left(u^{\beta}_{\mathrm{s,ref}}\right)^2} \quad \text{where} \quad \boldsymbol{u}^{\mathrm{s}}_{\mathrm{s,ref}} := \begin{pmatrix} u^{\alpha}_{\mathrm{s,ref}} \\ u^{\beta}_{\mathrm{s,ref}} \end{pmatrix} := \boldsymbol{T}_{\mathrm{p}}(\phi_k) \underbrace{\left(\boldsymbol{u}^k_{\mathrm{s,PI}} - \boldsymbol{u}^k_{\mathrm{s,comp}}\right)}_{=:\boldsymbol{u}^k_{\mathrm{s,ref}}}, \tag{13}$$

exceeds the maximally available voltage amplitude $\hat{u}_{\max}$ (in V), given by

$$\hat{u}_{\max}(u_{\mathrm{dc}}, \theta') := \underbrace{\frac{\sqrt{3}}{\sin(\theta') + \sqrt{3}\cos(\theta')}}_{\frac{\sqrt{3}}{2} \leq \,\cdot\, \leq 1} \cdot \tfrac{2}{3} u_{\mathrm{dc}} \leq \tfrac{2}{3} u_{\mathrm{dc}} \quad \text{where} \quad \theta' := \mathrm{mod}\Big(\underbrace{\mathrm{atan2}(u^{\beta}_{\mathrm{s,ref}}, u^{\alpha}_{\mathrm{s,ref}})}_{=:\theta}, \tfrac{\pi}{3}\Big) \in \big[0, \tfrac{\pi}{3}\big), \tag{14}$$

within the full (fault-free) voltage hexagon (see Fig. 2(b)). The maximally available amplitude $\hat{u}_{\max}$ of the converter depends on the voltage reference angle $\theta$ (in rad) and the DC-link voltage. Note that $\hat{u}_{\max}$ varies inside the voltage hexagon. It is larger for $\theta = 0^\circ$ or $\theta = 60^\circ$ (maximum voltage amplitude $\hat{u}_{\max}(u_{\mathrm{dc}}, 0) = \frac{2}{3} u_{\mathrm{dc}}$) than for $\theta = 30^\circ$ (minimum voltage amplitude $\hat{u}_{\max}(u_{\mathrm{dc}}, \pi/6) = \frac{1}{\sqrt{3}} u_{\mathrm{dc}}$). Invoking trigonometric identities leads to the expression of $\hat{u}_{\max}$ in (14) by only considering the first sector of the voltage hexagon, i.e. $\theta = \theta' \in \big[0^\circ, 60^\circ\big)$. Note that, in the first sector, $\theta'$ coincides with $\theta$. Then, by using the auxiliary phase angle $\theta'$ as defined in (14), the generalized formula for the available amplitude $\hat{u}_{\max}$ is obtained for all other sectors of the voltage hexagon.

*2) Cross-coupling feedforward compensation:* The compensation of cross-coupling terms in the current dynamics (5) is realized by the following feedforward control action

$$\boldsymbol{u}^k_{\mathrm{s,comp}}(t) := \begin{pmatrix} u^d_{\mathrm{s,comp}}(t) \\ u^q_{\mathrm{s,comp}}(t) \end{pmatrix} := -\omega_k(t) L_{\mathrm{s}} \boldsymbol{J} \boldsymbol{i}^k_{\mathrm{s}}(t) - \omega_k(t) \hat{\psi}_{\mathrm{pm}} \begin{pmatrix} 0 \\ 1 \end{pmatrix} = \begin{pmatrix} \omega_k(t) L_{\mathrm{s}} i^q_{\mathrm{s}}(t) \\ -\omega_k(t) L_{\mathrm{s}} i^d_{\mathrm{s}}(t) - \omega_k(t) \hat{\psi}_{\mathrm{pm}} \end{pmatrix}, \tag{15}$$

which (at least in steady state) cancels out the influence of the $q$-terms on the $d$-current dynamics and vice versa (cf. (5) in Sect. II-A).

*3) Reference voltage saturation:* To avoid undesirable and infeasible output voltages of converter by applying physically infeasible voltage reference vectors, the computed reference voltage vector $\boldsymbol{u}^{\mathrm{s}}_{\mathrm{s,ref}}$ as in (13) is saturated if necessary as follows

$$\boldsymbol{u}^{\mathrm{s}}_{\mathrm{s,ref,sat}}(t) = \begin{cases} \boldsymbol{u}^{\mathrm{s}}_{\mathrm{s,ref}}(t) & \text{, if } \hat{u}_{\mathrm{s,ref}}(t) \leq \hat{u}_{\max}(t) \\ \hat{u}_{\max}(t) \begin{pmatrix} \cos(\theta(t)) \\ \sin(\theta(t)) \end{pmatrix} & \text{, else.} \end{cases} \tag{16}$$

Note that the reference voltage saturation does only alter the length of the voltage vector not its direction.

*4) Space-vector modulation:* To generate the switching sequence and, in particular, the switching vector $\boldsymbol{s}_{\mathrm{s}}^{abc}$ based on the saturated reference voltage $\boldsymbol{u}_{\mathrm{s,ref,sat}}^{\mathrm{s}}$, a symmetrical space vector modulation (SVM) is used in this paper (cf. [26, Chap. 14] or [24, Sect. 2.4.1]). The boundary (adjacent) space vectors of the respective sector and, usually, both zero vectors $\boldsymbol{u}_{000}$ and $\boldsymbol{u}_{111}$ are applied to approximate the reference voltage vector $\boldsymbol{u}_{\mathrm{s,ref,sat}}^{\mathrm{s}}$ over one switching period $T_{\mathrm{sw}} = 1/f_{\mathrm{sw}}$ (in s). To do so, $T_{\mathrm{sw}}$ is divided into three time intervals: $T_1$ for the first non-zero vector, $T_2$ for the second non-zero vector and $T_0$ for the zero vectors such that $T_1 + T_2 + T_0 = T_{\mathrm{sw}}$ (see Fig. 2(b) and Fig. 6(a)). If the voltage reference vector is not saturated (cf. Sect. III-A3), this will lead—depending on the implementation of the SVM—e.g. to a negative time for $T_0$, which will cause strange behaviour of the SVM.

*5) Control performance of standard control system under an open-switch fault in $S_1$ (phase $a$):* To have a measure to evaluate the control performance of the standard and the fault-tolerant control systems, the total harmonic distortion (THD) is used. The total harmonic distortion $\mathrm{THD}_{i_{\mathrm{s}}^a}$ (in %) of e.g. the phase current $i_{\mathrm{s}}^a$ can be computed as follows [27]

$$\mathrm{THD}_{i_{\mathrm{s}}^a} := \frac{\sqrt{\sum_{n=2}^{\infty}(I_n^a)^2}}{I_1^a} \geq 0, \tag{17}$$

where $I_1^a$ and $I_n^a$ (both in A) are the rms values[1] of the fundamental and the $n$-th harmonic current component, respectively. Figure 10(a) shows the control performance of the standard control system (as in Fig. 4) under an open-switch fault in $S_1$ (phase $a$). The upper subplot illustrates reference ($i_{\mathrm{s,ref}}^d$ & $i_{\mathrm{s,ref}}^q$) and actuals currents ($i_{\mathrm{s}}^d$ & $i_{\mathrm{s}}^q$) in the $(d,q)$-reference frame, whereas the lower subplot shows the phase currents $i_{\mathrm{s}}^a$, $i_{\mathrm{s}}^b$ and $i_{\mathrm{s}}^c$ over time. It can be clearly seen, that the current $i_{\mathrm{s}}^a$ of phase $a$ is sinusoidal for the negative half-wave, but non-sinusoidal (close to zero) for the positive half-wave. This deviation leads to non-constant (as usually expected) currents $i_{\mathrm{s}}^d$ and $i_{\mathrm{s}}^q$ which significantly differ from their respective reference values $i_{\mathrm{s,ref}}^d$ and $i_{\mathrm{s,ref}}^q$ for (almost) all time. The current $i_{\mathrm{s}}^q$ tends to zero during the non-existing positive half-wave of $i_{\mathrm{s}}^a$. Moreover, even for the negative (correct) half-wave of $i_{\mathrm{s}}^a$, the current $i_{\mathrm{s}}^q$ is not capable of tracking its reference $i_{\mathrm{s,ref}}^q$. In particular, the non-constant and nonlinearly oscillating evolution of $i_{\mathrm{s}}^q$ leads to noticeable torque ripples. The non-zero current $i_{\mathrm{s}}^d$ does not contribute to the torque (recall (6)) but increases copper losses in the machine. In conclusion, the standard control system performance is not acceptable and must be improved to allow for a safe and uninterrupted operation of the wind turbine system.

### B. Proposed fault-tolerant control system (modified field-oriented control)

As illustrated in Fig. 10(a), the standard control system performance is poor and not acceptable when an open-swich fault is present. Without altering the hardware or the principle control system, three (software) modification are proposed to improve the control performance of the wind turbine system under open-switch faults in one phase. The three modifications are: 1) Extension of the anti-windup strategy, 2) Modification of the SVM and 3) Torque ripple minimization by injecting an optimal $d$-current. Each modification is explained in detail and its positive effect on the control performance of the fault-tolerant control system is illustrated by simulation results. Later, in Sect. IV, these modifications are implemented on a laboratory test bench and their effectiveness is validated by measurements. The block diagram of the improved and fault-tolerant control system is depicted in Fig. 5. The modifications are highlighted in blue.

*1) Extension of anti-windup strategy:* To improve the control performance under faults, in a first step, the anti-windup strategy (12) of the PI current controllers (11) is modified. The overshoots in the $q$-current during the open-switch fault in $S_1$ (recall Fig. 10(a)) are—at least partly—due to windup of the integral control action of the PI controllers during the positive (almost zero) half-wave of $i_{\mathrm{s}}^a$. To avoid this windup, an additional condition considering the current direction (see Fig. 5) must be introduced which leads to the extended anti-windup decision function

$$f_{\mathrm{aw}}^{\star}\left(\hat{u}_{\mathrm{s,ref}}, i_{\mathrm{s}}^a\right) := \begin{cases} 1 & \text{, if } \hat{u}_{\mathrm{s,ref}} \leq \hat{u}_{\max}(u_{\mathrm{dc}}, \theta') \ \text{ AND } \ i_{\mathrm{s}}^a < \hat{\imath}_{\mathrm{aw}} < 0 \\ 0 & \text{, else} \end{cases} \tag{18}$$

for open-switch faults in $S_1$ (phase $a$), which replaces $f_{\mathrm{aw}}(\cdot)$ in (11). The constant $\hat{\imath}_{\mathrm{aw}} < 0$ (in A) represents the maximally admissible anti-windup current and should be chosen negative to account for the chattering of the phase current $i_{\mathrm{s}}^a$ around zero (recall Fig. 10(a)). Note that, for any other faulty phase with open-switch fault in $S_2$ (or $S_3$), the respective phase current direction of $i_{\mathrm{s}}^b$ (or $i_{\mathrm{s}}^c$) must be considered in (18) instead of $i_{\mathrm{s}}^a$.

In Fig. 10(b), the improved control system performance due to the extended anti-windup strategy (18) is shown. The $abc$-currents (lower subplot) do not alter much (almost no improvement is visible) and the THD reduces slightly to $\mathrm{THD}_{i_{\mathrm{s}}^a} = 41.4\,\%$. But the tracking performance of the currents $i_{\mathrm{s}}^d$ and, in particular, $i_{\mathrm{s}}^q$ is improved substantially. During the positive (almost zero) half-wave of $i_{\mathrm{s}}^a$, both currents still do not perfectly track their references; but during the negative half-wave of $i_{\mathrm{s}}^a$, both currents are capable of (almost) asymptotic reference tracking. Especially, the current ripple in the $q$-current is drastically reduced during the negative half-wave of $i_{\mathrm{s}}^a$.

[1] The root mean square (rms) value is defined by $I := \sqrt{\frac{1}{T}\int_{t-T}^{t} i(\tau)^2 \mathrm{d}\tau}$ with fundamental period $T = 1/f$ of the current $i(\cdot)$.

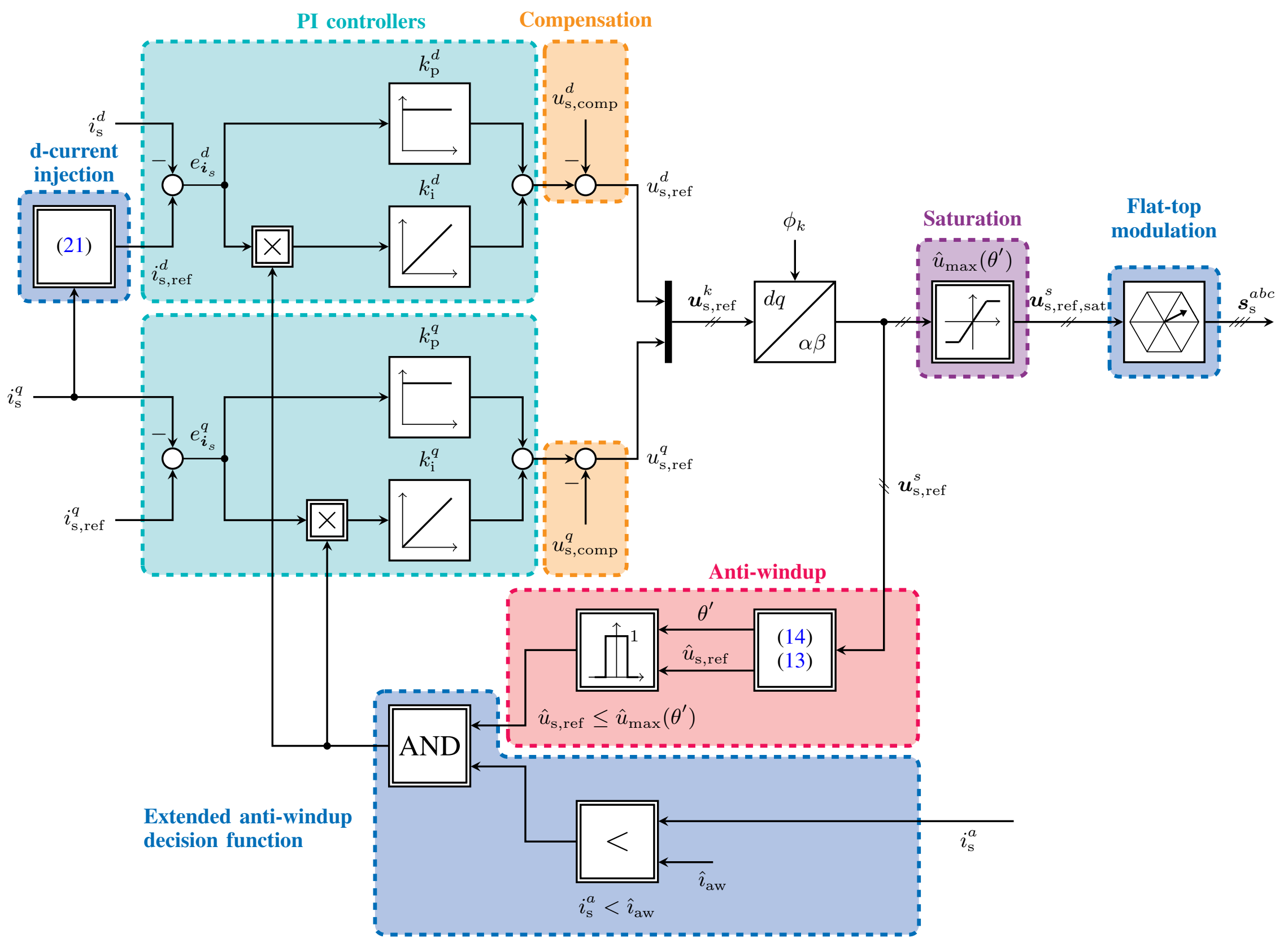

Fig. 5: *Fault-tolerant control system (modified field-oriented control; changes in blue) for an open-switch fault in $S_1$ (phase $a$): PI controllers with improved anti-windup, cross-coupling feedforward compensation and reference voltage saturation.*

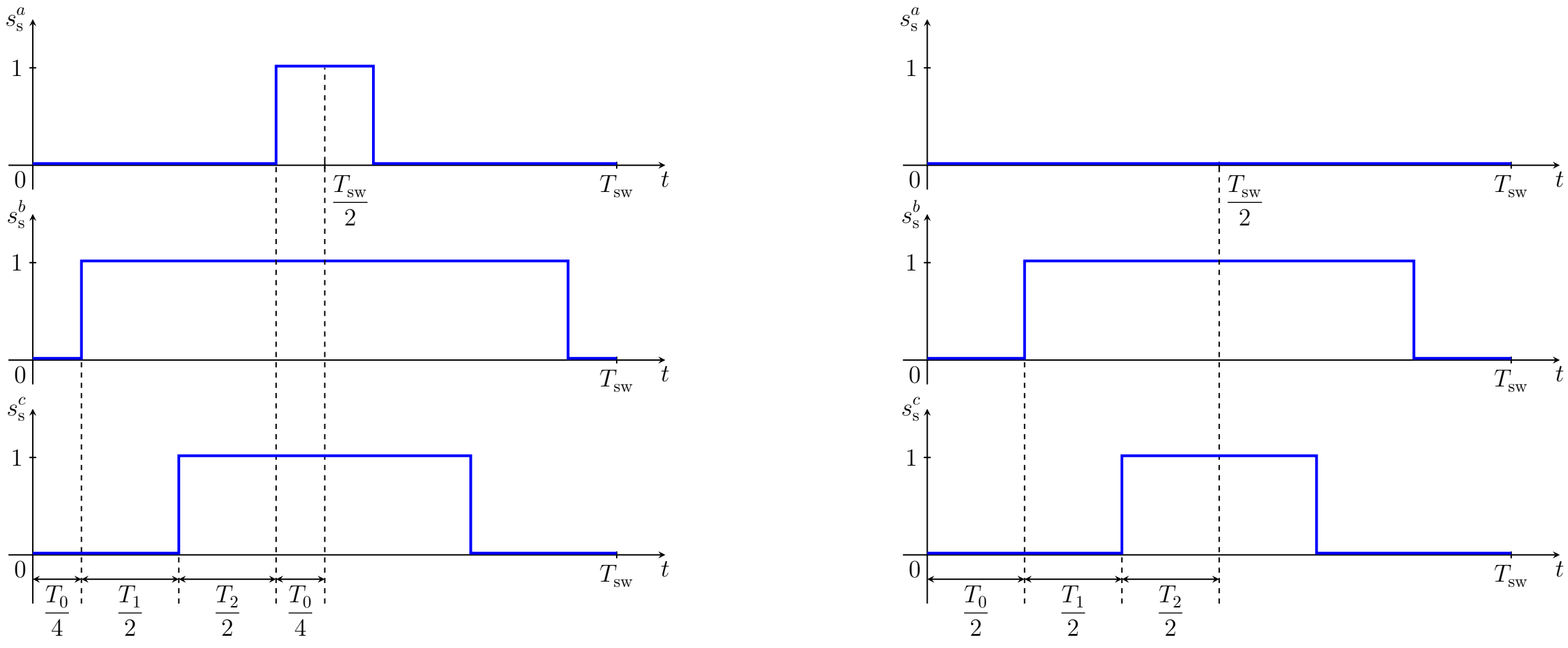

(a) Standard SVM. (b) Flat-top modulation.

Fig. 6: *Switching patterns to generate a voltage reference vector $\boldsymbol{u}^s_{s,\text{ref}}$ in sector III by a linear combination of the space vectors $\boldsymbol{u}^s_{010}$, $\boldsymbol{u}^s_{011}$ and* (a) *both zero vectors $\boldsymbol{u}^s_{000}$ and $\boldsymbol{u}^s_{111}$,* (b) *only the zero vector $\boldsymbol{u}^s_{000}$.*

*2) Modification of SVM (flat-top modulation):* The second step to improve the control performance during open-switch faults is the modification of the space-vector modulation. Note that any open-switch fault in one of the *upper* switches (i.e. $S_1$, $S_2$ or $S_3$) leads to a shifted zero vector $\boldsymbol{u}^s_{111}$ (for $i^a_s \geq 0$, see Fig. 2(b)); whereas any open-switch fault in one of the *lower* switches (i.e. $\overline{S}_1$, $\overline{S}_2$ or $\overline{S}_3$) shifts the other zero vector $\boldsymbol{u}^s_{000}$. Hence, one of the zero vectors is not zero any more. Then, using *flat-top* modulation allows to use only the non-shifted zero vector (cf. [26, Chap. 14.6], [28]). For example, for an open-switch fault in $S_1$, $S_2$ or $S_3$, only the zero vector $\boldsymbol{u}^s_{000}$ is applied (see Fig. 6(b)). For faults in $\overline{S}_1$, $\overline{S}_3$ or $\overline{S}_2$ only the zero vector $\boldsymbol{u}^s_{111}$ is used. In Fig. 10(c) the positive effect of the flat-top modulation on the control performance and the THD of $i^a_s$ is illustrated.

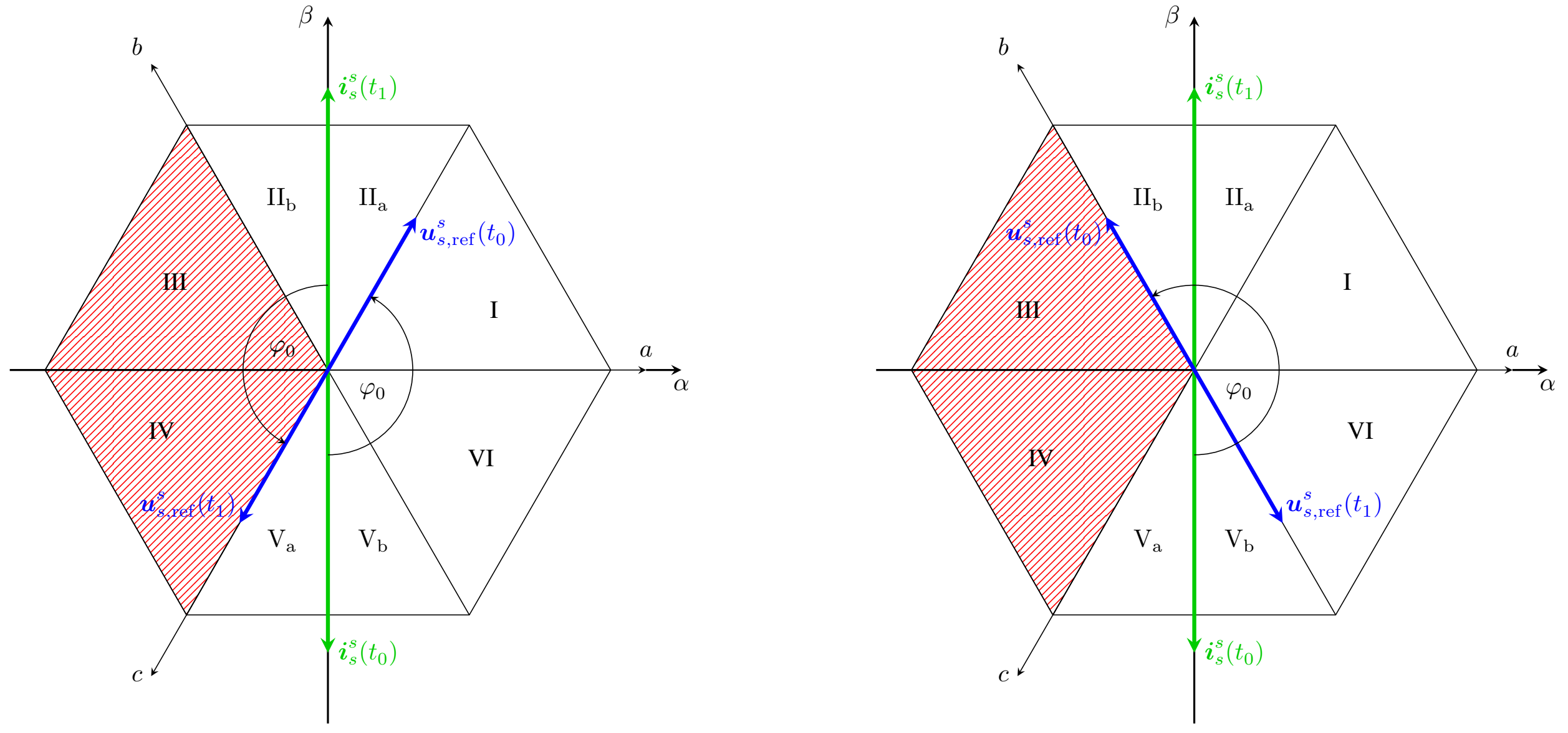

(a) $\varphi_0 = 150^\circ$. (b) $\varphi_0 = 210^\circ$.

Fig. 7: *Stator current vector* $\boldsymbol{i}_\mathrm{s}^s$ *and voltage reference vector* $\boldsymbol{u}_{\mathrm{s,ref}}^s$ *at time* $t_0$ *and* $t_1$ *for* (a) $\varphi_0 = 150^\circ$ *and* (b) $\varphi_0 = 210^\circ$.

The upper subplots show the $d$ and $q$ currents and their references, whereas the lower subplot depicts the $abc$-currents for the modified control system with extended anti-windup and modified SVM. Clearly, the intervals where $i_\mathrm{s}^a \approx 0\,\mathrm{A}$ are significantly shorter. Moreover, positive and almost sinusoidal $i_\mathrm{s}^a$ currents are feasible again due to the modified SVM. So, the THD value is drastically reduced to $\mathrm{THD}_{i_\mathrm{s}^a} = 19.5\,\%$ and the tracking control performances of the currents $i_\mathrm{s}^d$ and $i_\mathrm{s}^q$ are improved as well.

*3) Injection of optimal $d$-current:* The last improvement is to inject an optimal (additional) $d$-current to minimize the THD of the faulty phase even further. In the following, an open-switch fault in $S_1$ is considered. For other open-switch faults, the modifications are straight forward. As discussed in Sect. II-B2, for a fault in $S_1$ and $i_\mathrm{s}^a \geq 0$, the output voltages that can be provided by the faulty converter are limited. For e.g. $i_\mathrm{s}^a > 0$, only voltage vectors from the sectors III and IV are feasible (see Fig. 3). The principle idea of the optimal $d$-current injection is to generate auxiliary reference voltage vectors within those two feasible sectors as long as possible. The goal is to determine an optimal *phase shift* $\varphi_0$ (in rad or $^\circ$) between stator current $\boldsymbol{i}_\mathrm{s}^s$ and reference voltage $\boldsymbol{u}_{\mathrm{s,ref}}^s$.

To illustrate the idea, in Fig. 7, the phase shifts $\varphi_0 = 150^\circ$ and $\varphi_0 = 210^\circ$ are shown for two time instants $t_0$ and $t_1$ where $\boldsymbol{i}_\mathrm{s}^s(t_0)$ and $\boldsymbol{i}_\mathrm{s}^s(t_1)$ are located on the negative and positive $\beta$-axis, respectively. Note that, if the phase current $i_\mathrm{s}^a$ is non-negative, the stator current *space vector* $\boldsymbol{i}_\mathrm{s}^s$ is located in the *right* half-plane (see Fig. 7). More precisely, at $t_0$ with $i_\mathrm{s}^a(t_0) = 0$ ($i_\mathrm{s}^a$ becomes positive thereafter), $\boldsymbol{i}_\mathrm{s}^s$ lies on the negative $\beta$-axis; whereas, at $t_1$ with $i_\mathrm{s}^a(t_1) = 0$ ($i_\mathrm{s}^a$ becomes negative afterwards), $\boldsymbol{i}_\mathrm{s}^s$ is aligned with the positive $\beta$-axis. Clearly, within the interval $[t_0, t_1]$, the current moves by $180^\circ$ and, optimally, the corresponding stator voltage reference *space vector* $\boldsymbol{u}_{\mathrm{s,ref}}^s$ should be within the sectors III and IV as long as possible in order to apply feasible and correct voltages to the generator. However, as these two sectors represent only span over $120^\circ$, it is not possible to apply the correct voltages during the whole non-negative half-wave of $i_\mathrm{s}^a$ during an open-switch fault in $S_1$. During the remaining $60^\circ$, incorrect voltages will be applied by the faulty converter which affect the shape of the currents and cause deviations from the desired sinusoidal waveform.

Depending on the phase shift $\varphi_0$ between stator current and reference voltage, different parts of the non-negative half-wave of $i_\mathrm{s}^a$ are affected by the fault. For $\varphi_0 = 150^\circ$ (see Fig. 7(a)), $\boldsymbol{u}_{\mathrm{s,ref}}^s$ starts in sector II at time $t_0$. So, for the first $60^\circ$ of the current half-wave, incorrect voltages are applied to the generator. As soon as $\boldsymbol{u}_{\mathrm{s,ref}}^s$ enters sector III, the correct voltages can be provided (even) by the faulty converter. When $\boldsymbol{u}_{\mathrm{s,ref}}^s$ is in the sectors III and IV, the correct voltages give rise to a sinusoidal current. For $\varphi_0 = 210^\circ$ (see Fig. 7(b)), the behaviour is flipped: At time $t_0$, $\boldsymbol{u}_{\mathrm{s,ref}}^s$ starts already in the feasible sector III and, hence, during the first $120^\circ$ of the non-negative current half-wave, the correct voltages are applied. But, as soon as $\boldsymbol{u}_{\mathrm{s,ref}}^s$ enters sector V, incorrect voltages are generated by the converter for the rest of the half-wave until time $t_1$. Concluding, in order to fully benefit from the two feasible voltage sectors III and IV, where the correct voltages can be generated for $i_\mathrm{s}^a > 0$, the phase shift must be within the interval $\varphi_0 \in \left[150^\circ, 210^\circ\right]$. The observations above are also validated by the simulation results presented in Fig. 8: For $\varphi_0 = 150^\circ$ (see Fig. 8(a)), during the first $60^\circ$, the phase current $i_\mathrm{s}^a$ jitters around zero. In contrast to the last $120^\circ$, where the desired sinusoidal characteristic is achieved. For $\varphi_0 = 210^\circ$ (see Fig. 8(b)), the phase current $i_\mathrm{s}^a$ exhibits a sinusoidal characteristic during the first $120^\circ$, whereas, for the last $60^\circ$, it is deteriorated.

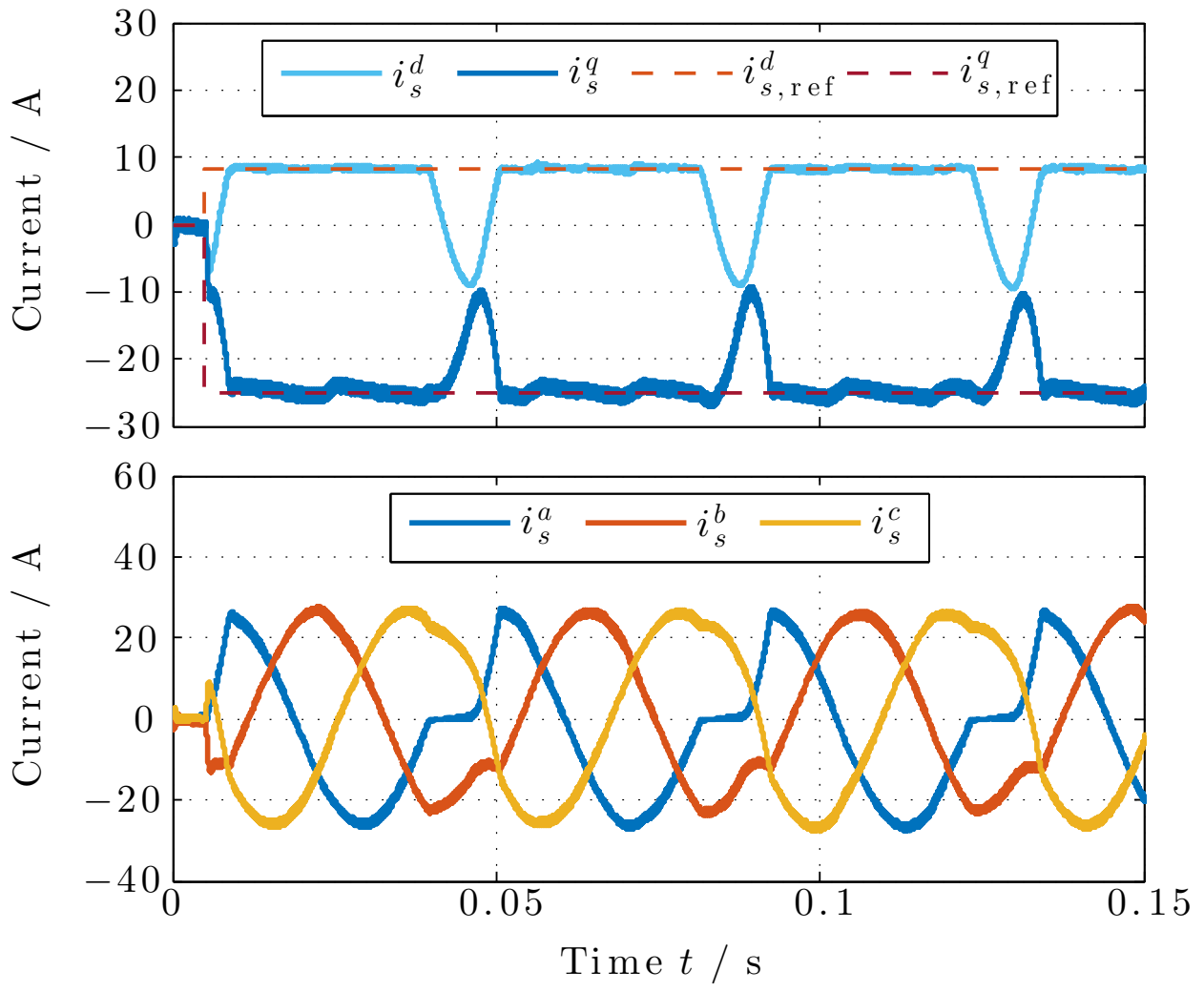

(a) $\varphi_0 = 150^\circ$: $\mathrm{THD}_{i_\mathrm{s}^a} = 31.4\,\%$.

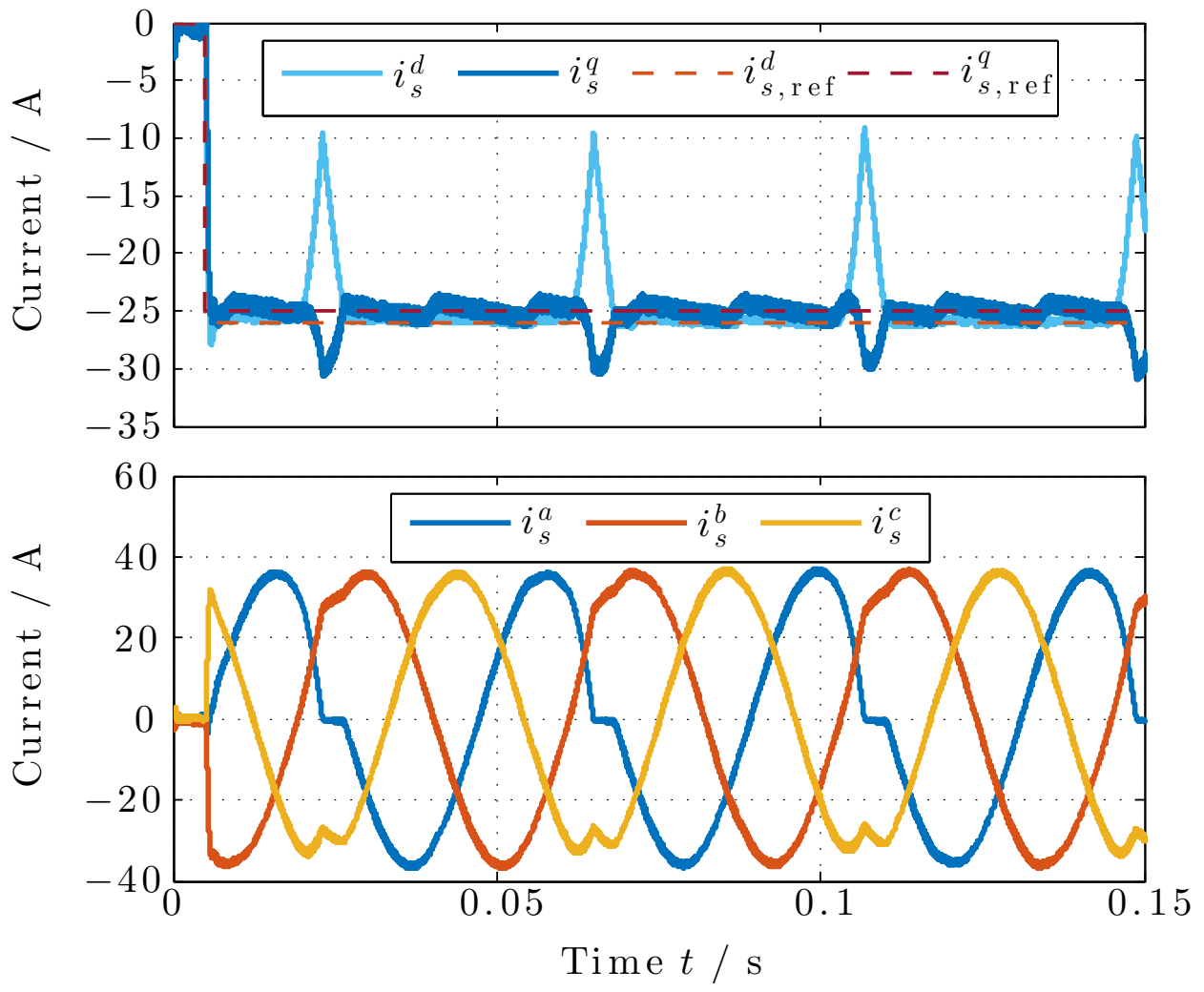

(b) $\varphi_0 = 210^\circ$: $\mathrm{THD}_{i_\mathrm{s}^a} = 12.2\,\%$.

Fig. 8: *Currents $\boldsymbol{i}_\mathrm{s}^k$ (with references) and $\boldsymbol{i}_\mathrm{s}^{abc}$ for different phase shifts:* (a) $\varphi_0 = 150^\circ$ *and* (b) $\varphi_0 = 210^\circ$.

**Remark III.1.** *For a converter outputting power—e.g. for PMSMs in motor mode or for grid-side inverters in wind turbine systems, solely phase shifts of $\varphi_0 \in \big(-90^\circ, 90^\circ\big)$ are feasible. Hence, it is not possible to benefit from the sectors III and IV.*

The phase shift $\varphi_0$ can be altered by the injection of a $d$-current which also changes the ratio between active power $p$ (in W) and reactive power $q$ (in var), since

$$q = p \cdot \tan(\varphi_0) \quad \text{where} \quad p = \tfrac{3}{2}(\boldsymbol{u}_\mathrm{s}^k)^\top \boldsymbol{i}_\mathrm{s}^k \quad \text{and} \quad q = \tfrac{3}{2}(\boldsymbol{u}_\mathrm{s}^k)^\top \boldsymbol{J}\boldsymbol{i}_\mathrm{s}^k \qquad \text{(see e.g. [25]).} \tag{19}$$

Moreover, note that, due to (6), $i_\mathrm{s}^d$ can be chosen independently of the desired torque. In order to derive an analytical expression for the reference current $i_\mathrm{s,ref}^d$ of the to-be-injected current $i_\mathrm{s}^d$, the following assumption is imposed:

**Assumption (A.1)** *The copper losses in the PMSM are negligible and its current dynamics are in steady state, i.e.*

$$R_\mathrm{s} \approx 0\,\Omega \quad \textit{and} \quad \tfrac{\mathrm{d}}{\mathrm{d}t}\boldsymbol{i}_\mathrm{s}^k = \boldsymbol{0}_2. \tag{20}$$

Solving (5) (in steady state) for $\boldsymbol{u}_\mathrm{s}^k$ and inserting the result into (19) gives

$$\underbrace{n_\mathrm{p}\omega_\mathrm{m}}_{\omega_k}\Big[L_\mathrm{s}(i_\mathrm{s}^d)^2 + L_\mathrm{s}(i_\mathrm{s}^q)^2 + \hat{\psi}_\mathrm{pm} i_\mathrm{s}^d\Big] = \Big[R_\mathrm{s}(i_\mathrm{s}^d)^2 + R_\mathrm{s}(i_\mathrm{s}^q)^2 + n_\mathrm{p}\omega_\mathrm{m}\hat{\psi}_\mathrm{pm} i_\mathrm{s}^q\Big]\tan(\varphi_0),$$

which is a second-order polynomial in $i_\mathrm{s}^d$. Its root with the smaller amplitude is used as $d$-current reference[2], i.e.

$$\begin{aligned}
i_\mathrm{s,ref}^d(i_\mathrm{s}^q) &= -\frac{n_\mathrm{p}\omega_\mathrm{m}\hat{\psi}_\mathrm{pm}}{2\big(n_\mathrm{p}\omega_\mathrm{m}L_\mathrm{s} - R_\mathrm{s}\tan(\varphi_0)\big)} + \sqrt{\frac{(n_\mathrm{p}\omega_\mathrm{m}\hat{\psi}_\mathrm{pm})^2}{4\big(n_\mathrm{p}\omega_\mathrm{m}L_\mathrm{s} - R_\mathrm{s}\tan(\varphi_0)\big)^2} - (i_\mathrm{s}^q)^2 + \frac{n_\mathrm{p}\omega_\mathrm{m}\hat{\psi}_\mathrm{pm}\, i_\mathrm{s}^q\, \tan(\varphi_0)}{2\big(n_\mathrm{p}\omega_\mathrm{m}L_\mathrm{s} - R_\mathrm{s}\tan(\varphi_0)\big)}} \\
&= -\frac{n_\mathrm{p}\omega_\mathrm{m}\hat{\psi}_\mathrm{pm}}{2\big(n_\mathrm{p}\omega_\mathrm{m}L_\mathrm{s} - R_\mathrm{s}\tan(\varphi_0)\big)} + \sqrt{\frac{(n_\mathrm{p}\omega_\mathrm{m}\hat{\psi}_\mathrm{pm})^2}{4\big(n_\mathrm{p}\omega_\mathrm{m}L_\mathrm{s} - R_\mathrm{s}\tan(\varphi_0)\big)^2} - (i_\mathrm{s}^q)^2 + \frac{n_\mathrm{p}\omega_\mathrm{m}\hat{\psi}_\mathrm{pm}\, i_\mathrm{s}^q\, \tan(\varphi_0)}{1\big(n_\mathrm{p}\omega_\mathrm{m}L_\mathrm{s} - R_\mathrm{s}\tan(\varphi_0)\big)}} \\
&\overset{[R_\mathrm{s}\approx 0]}{=} -\frac{\hat{\psi}_\mathrm{pm}}{2\,L_\mathrm{s}} + \sqrt{\Big(\frac{\hat{\psi}_\mathrm{pm}}{2\,L_\mathrm{s}}\Big)^2 - (i_\mathrm{s}^q)^2 + \frac{\hat{\psi}_\mathrm{pm}}{L_\mathrm{s}}\, i_\mathrm{s}^q\, \tan(\varphi_0)}.
\end{aligned} \tag{21}$$

Hence, for large machines with $R_\mathrm{s} \approx 0$, the reference current $i_\mathrm{s,ref}^d$ depends on the machine parameters $L_\mathrm{s}$ & $\hat{\psi}_\mathrm{pm}$, the current $i_\mathrm{s}^q$ (or its reference $i_\mathrm{s,ref}^q$) and the desired phase angle $\varphi_0$. There exists an optimal value for $\varphi_0$ to minimize the THD of the phase current $i_\mathrm{s}^a$. For the considered machine, the optimal value $\boxed{\varphi_{0,\mathrm{opt}} = \varphi_0 = 197^\circ}$ was found by iterative simulations: These results are depicted in Fig. 9(a). Clearly, for other machines, the optimal value might be different.

Finally, in Fig. 10(d), the simulation results for the overall fault-tolerant control system with extended anti-windup, modified SVM (flat-top modulation) *and* optimally injected $d$-current are shown. The upper subplot depicts the currents $i_\mathrm{s}^d$ and $i_\mathrm{s}^q$ and their reference values, whereas the lower subplot illustrates the shape of the $abc$-currents. The THD value for this scenario is $\mathrm{THD}_{i_\mathrm{s}^a} = 9.4\,\%$, which is clearly the lowest compared to the other simulation results in Figures 10(a), 10(b) and 10(c).

[2]The solution with $-$ in front of the root would lead to a higher current magnitude.

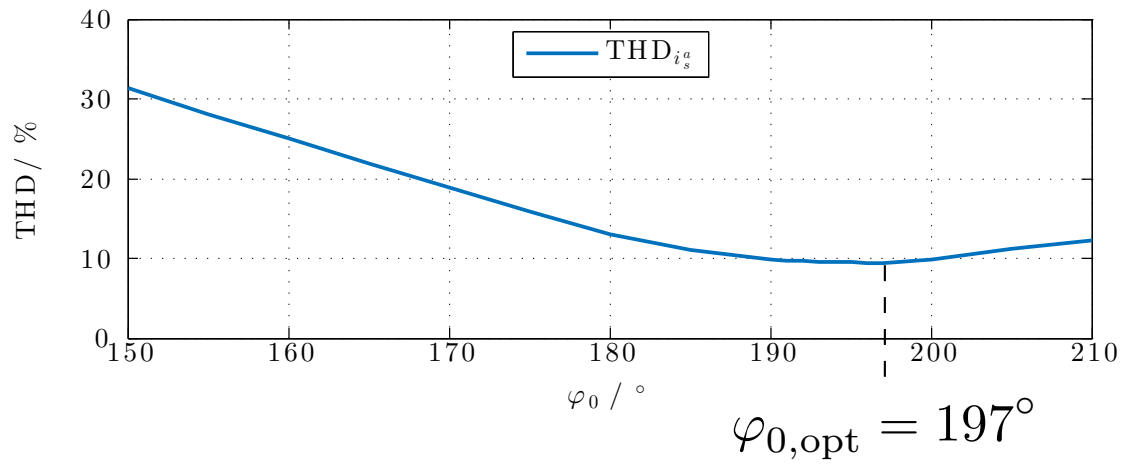

(a) THD$_{i_s^a}$ for different phase angles $\varphi_0$.

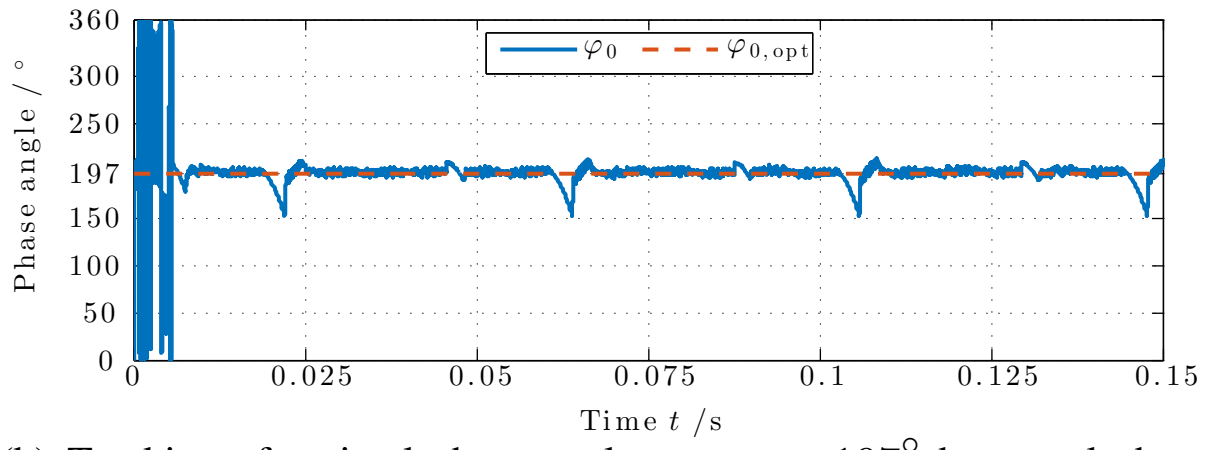

(b) Tracking of optimal phase angle $\varphi_{0,\mathrm{opt}} = 197°$ by actual phase angle $\varphi_0$.

Fig. 9: THD$_{i_s^a}$ *of phase current* $i_s^a$ *for different phase angles* $\varphi_0 \in [150°, 210°]$ *and tracking of optimal phase angle* $\varphi_{0,\mathrm{opt}} = 197°$ *under open-switch fault in* $S_1$ *(phase* $a$*).*

Moreover, the reference tracking capability, in particular, of $i_s^q$ is the best which implies that the torque ripples[3] are also minimized (recall (6)) leading to less stress on the mechanical drive train. In Fig. 9(b), the tracking of the optimal phase angle $\varphi_{0,\mathrm{opt}} = 197°$ by the actual phase angle $\varphi_0$ is shown. Due to the remaining time periods, where the correct voltage vector cannot be applied and the phase current is close to zero, the optimal phase angle cannot be achieved for all time.

**Remark III.2** (Possible limitation of the $d$-current injection in wind turbine systems)**.** *Due to the current rating/limitation of the machine-side converter, the additional injection of a* $d$*-current might not be feasible up to the rated torque of the isotropic generator in wind turbine systems. Therefore, the uninterrupted operation of the wind turbine system under open-switch faults might not be possible for all wind speeds unless the pitch control system is incorporated into the fault-tolerant control system. The turbine torque (proportional to the wind speed and the pitch angle) must be decreased by changing the pitch angle such that the current rating of the machine-side converter is not exceeded (for details see [25], [30, Chap. 8]).*

## IV. Implementation and Experimental Verification

In this section, implementation, experimental validation and comparison of simulation and measurement results are discussed. Three *experiments* in the laboratory are conducted to

($E_1$) validate the accuracy of the proposed mathematical model (8) of the converter with open-switch fault (in $S_1$) against a real electrical drive system with open-switch fault;
($E_2$) verify the effectiveness of the proposed modifications (such as extension of the anti-windup strategy, flat-top modulation and injection of an optimal $d$-current) on the generator control performance and to compare simulation and experimental results; and
($E_3$) illustrate the impact of an occurring open-switch fault (after a fault-free interval) and then, step-by-step, the positive effect of each proposed modification on the control performance of the laboratory electrical drive system.

### A. Experimental setup and implementation

The measurements where conducted on a 10 kW laboratory test bench as depicted in Fig. 11. The anisotropic *reluctance synchronous machine* (RSM) is speed-controlled (with underlying nonlinear current controllers [31]). The isotropic *permanent-magnet synchronous machine* is used as generator and current-controlled as described in the previous sections. Both drives are controlled by a dSPACE real-time system which applies the switching signals (switching vectors) to the respective inverter/converter. Both converters are connected back-to-back. The PMSM converter is modified such that each upper and lower switch can be addressed individually and allows to emulate open-switch faults. For all experiments, without loss of generality, open-switch faults in $S_1$ (phase $a$) were considered, simulated and emulated.

The implementation for simulations and measurements was performed using Matlab/Simulink. In Fig. 12, the block diagram of the implementation is shown. The parameters of the laboratory setup are listed in Tab. II (where $\Theta = \Theta_{\mathrm{RSM}} + \Theta_{\mathrm{PMSM}}$) and coincide with those used for the simulations. Note that the measured currents were filtered (by an analogue filter in the converter) and, then, sampled with the switching frequency; whereas the simulated currents were *not* filtered.

### B. Discussion of experiments

*1) Experiment ($E_1$):* The simulation *and* measurement scenario of this experiment is as follows: The PMSM-side converter emulates an open-switch fault in $S_1$ and the standard control system (as described in Sect. III-A) was implemented. Simulation

[3]Note that if the produced generator torque in wind turbine systems does not equal its reference value, wind turbine efficiency and power production are reduced [29].

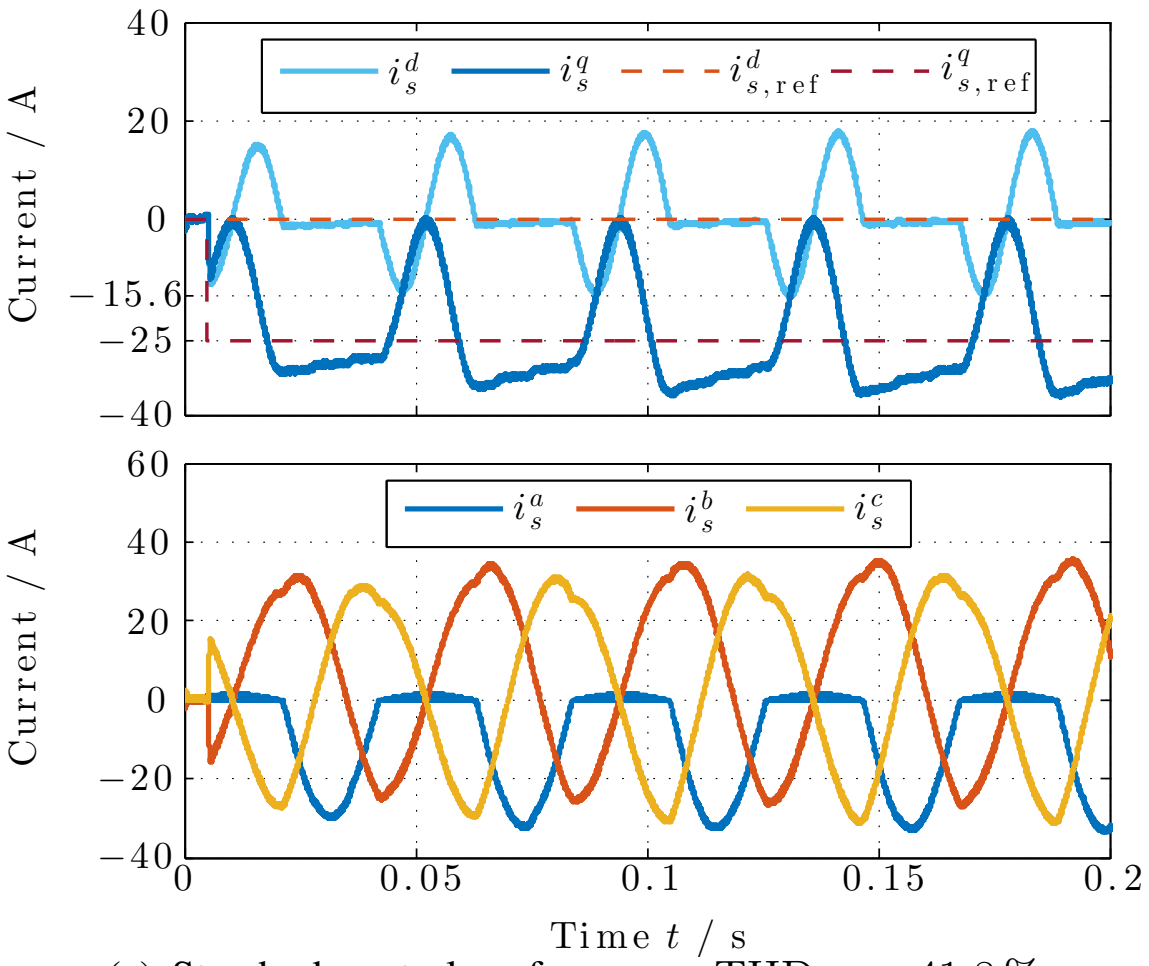

(a) Standard control performance: $\mathrm{THD}_{i_s^a} = 41.8\,\%$.

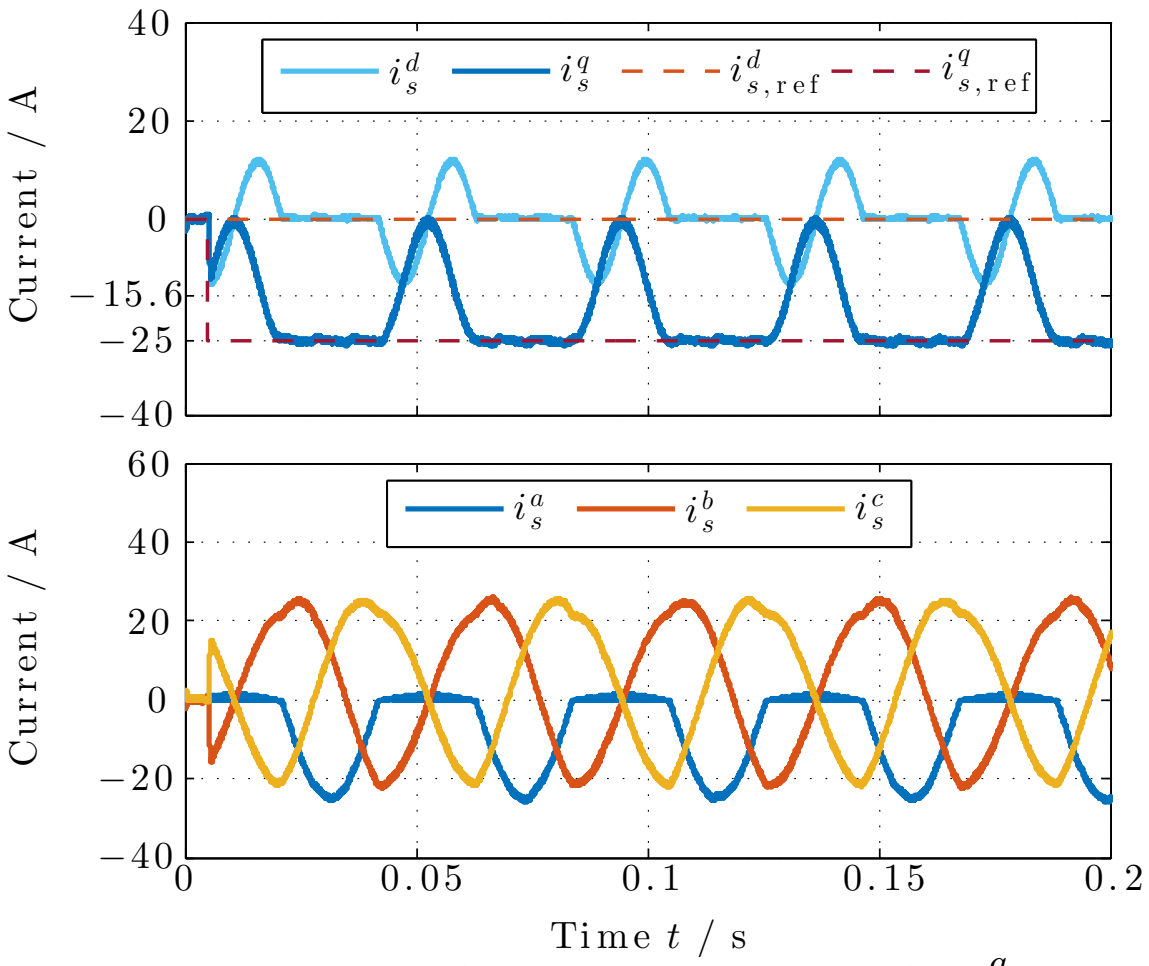

(b) Improved control performance (in particular for $i_s^q$) due to extended anti-windup (18): $\mathrm{THD}_{i_s^a} = 41.4\,\%$.

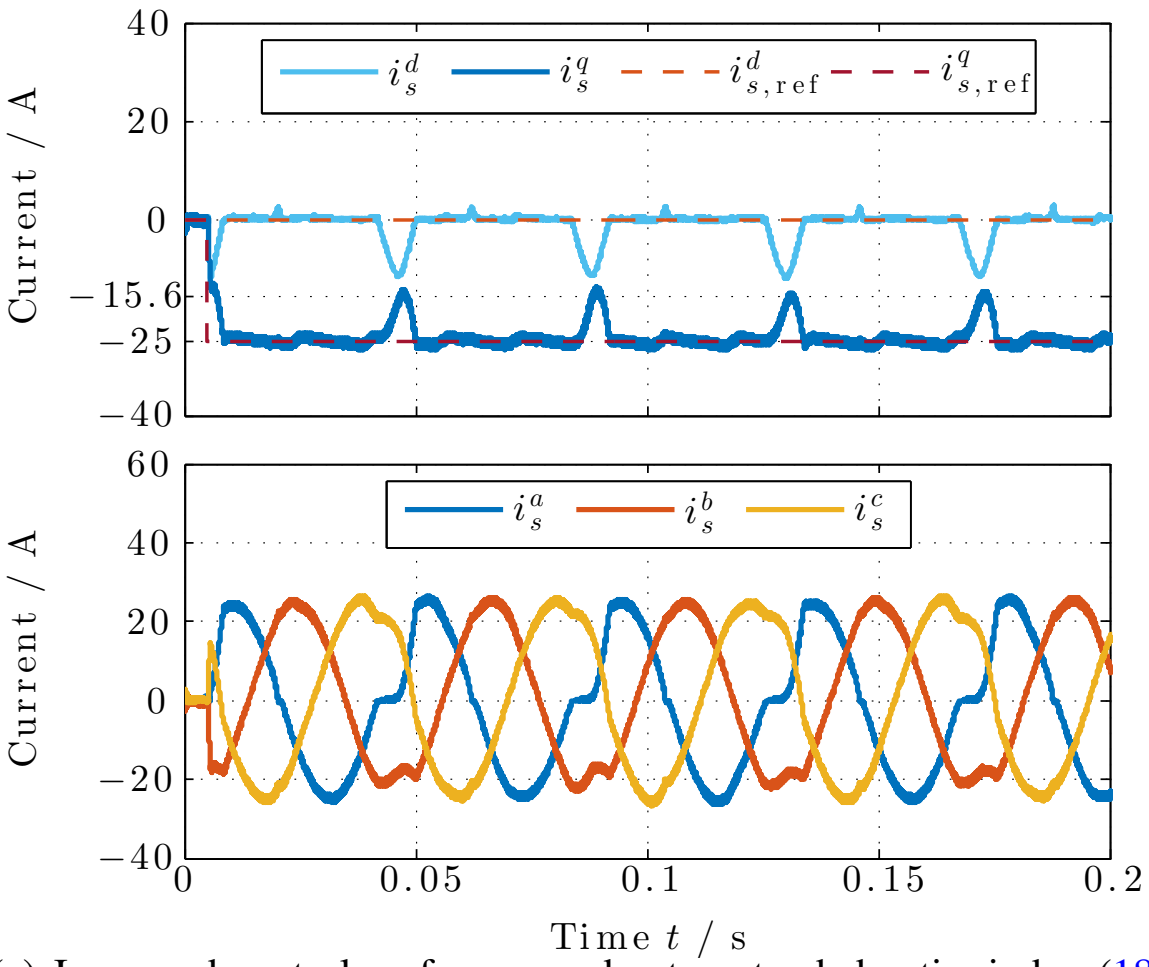

(c) Improved control performance due to extended anti-windup (18) and modified SVM (flat-top modulation): $\mathrm{THD}_{i_s^a} = 19.5\,\%$.

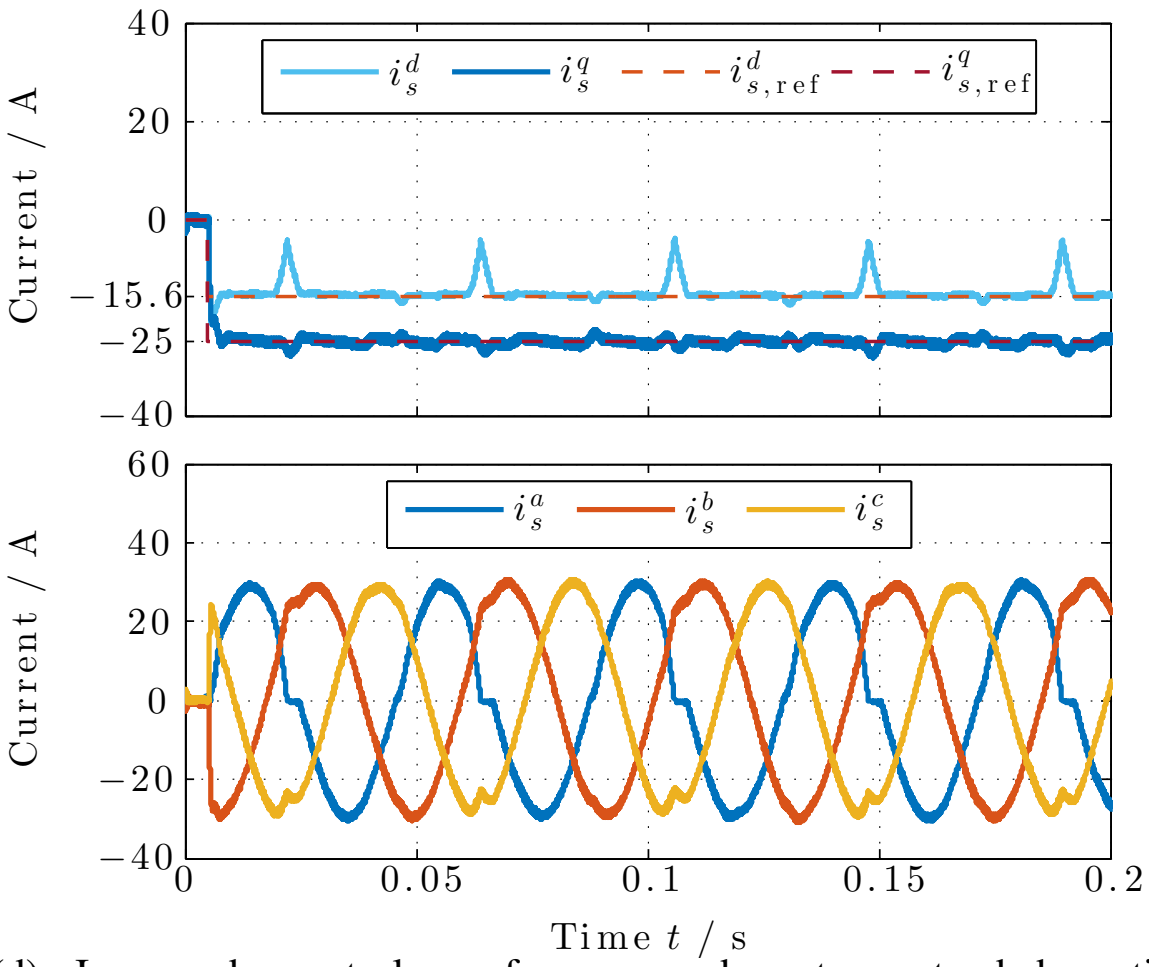

(d) Improved control performance due to extended anti-windup (18), modified SVM (flat-top modulation) and $d$-current injection: $\mathrm{THD}_{i_s^a} = 9.4\,\%$.

Fig. 10: *Comparative control system performance under open-switch fault in $S_1$ (phase $a$).*

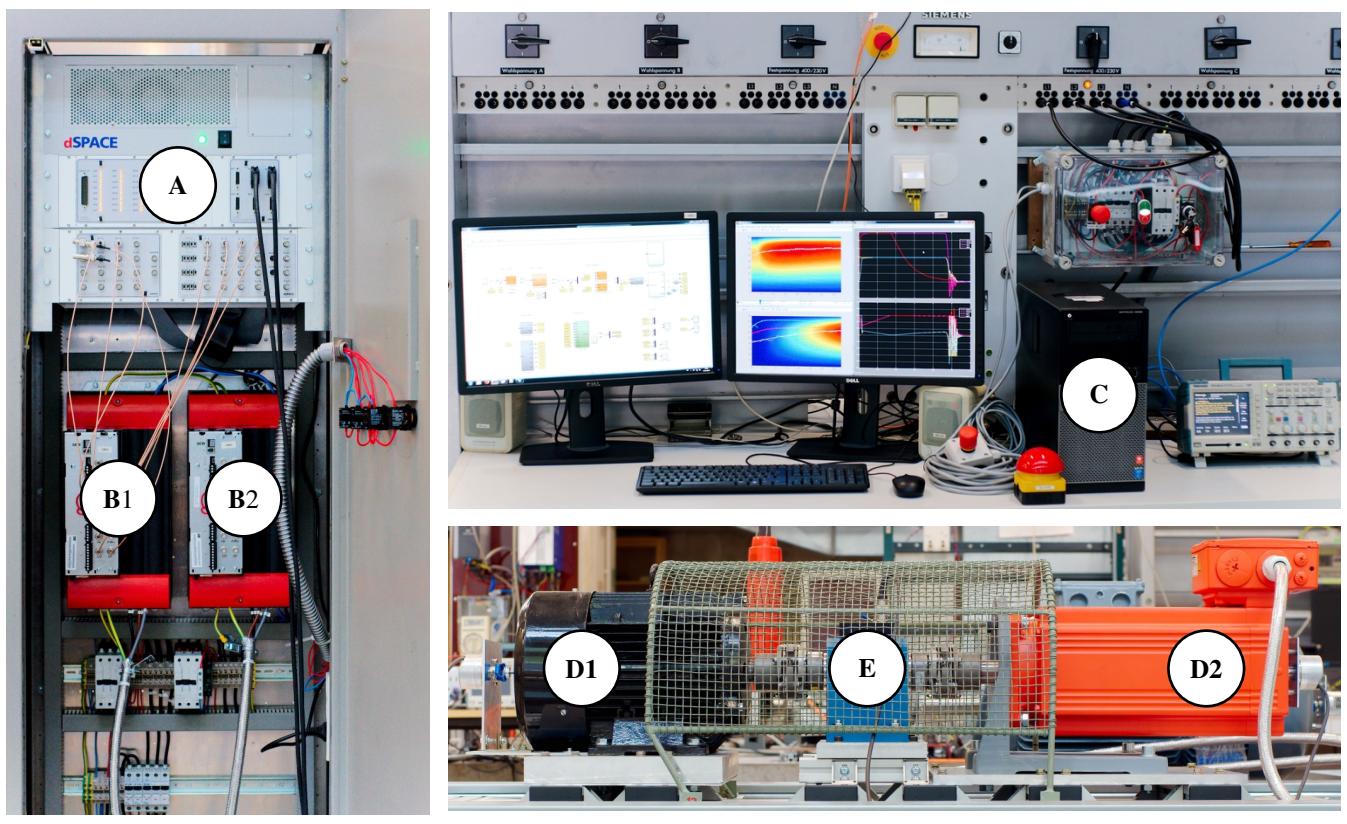

Fig. 11: *Laboratory test bench with dSPACE real-time system (A), voltage-source inverters (B1) and (B2) connected back-to-back, Host-PC (C), reluctance synchronous machine (RSM; D1) and permanent-magnet synchronous machine (PMSM; D2), and torque sensor (E).*

TABLE II: *Simulation and measurement data.*

| Description | Symbol | Value and unit |
|---|---|---|
| *Simulation parameters* | | |
| ODE-Solver (fixed-step) | | Runge-Kutta (ode4) |
| sampling time | $h$ | $1\,\mu s$ |
| *Converter* | | |
| DC-link voltage | $u_{dc}$ | $565\,V$ |
| switching frequency | $f_{sw}$ | $8\,kHz$ |
| *Permanent-magnet synchronous machine (generator, isotropic)* | | |
| stator resistance | $R_s$ | $0.11\,\Omega$ |
| stator inductance | $L_s$ | $3.35\,mH$ |
| PM-flux linkage | $\hat{\psi}_{pm}$ | $0.377\,V\,s$ |
| number of pole pairs | $n_p$ | 3 |
| machine inertia | $\Theta_{PMSM}$ | $163 \cdot 10^{-4}\,kg\,m^2$ |
| *Reluctance synchronous machine (anisotropic)* | | |
| stator resistance | $R_s$ | $0.4\,\Omega$ |
| stator inductances | $L_s^d \neq L_s^q$ | nonlinear (see [32, Fig. 2]) |
| number of pole pairs | $n_p$ | 2 |
| machine inertia | $\Theta_{RSM}$ | $189 \cdot 10^{-4}\,kg\,m^2$ |
| *Current control system of PMSM* | | |
| PI controller gains | $k_p^d = k_p^q$ | $8.93\,\frac{V}{A}$ |
| | $k_i^d = k_i^q$ | $293.3\,\frac{V}{A\,s}$ |
| maximum anti-windup current | $\hat{\imath}_{aw}$ | $-1\,A$ |

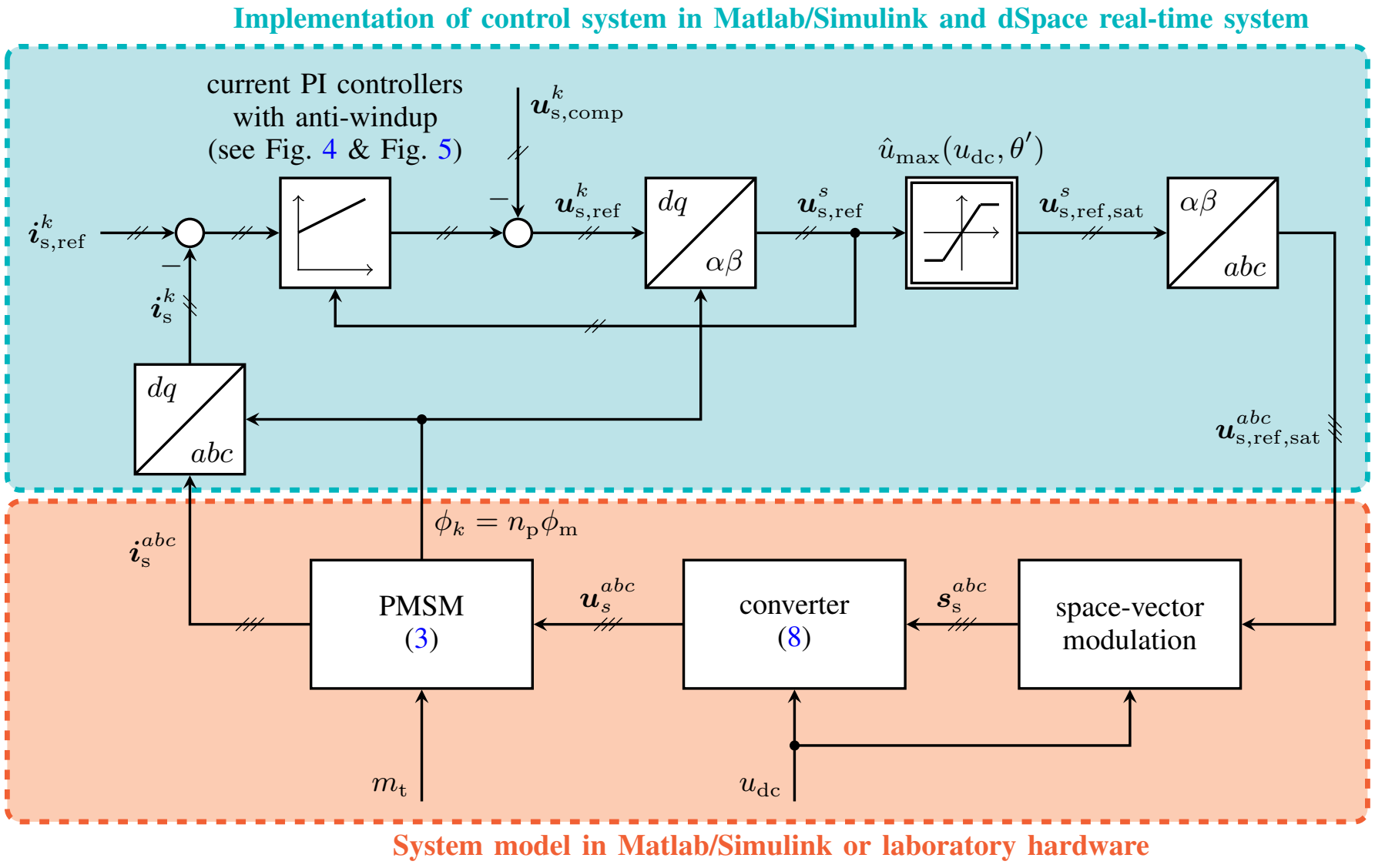

Fig. 12: *Block diagram of the implementation of SVM, converter with open-switch fault in $S_1$, PMSM and current control system in Matlab/Simulink and on the dSPACE real-time system.*

and measurement results of Experiment ($E_1$) are shown in Fig. 13(a). The measured quantities are labeled with the additional subscript "meas". Obviously, simulation and measurement results match very closely. Hence, the proposed mathematical model (8) is valid and allows to simulate the behavior of the real system precisely. Note that, due to the smaller power rating of the RSM, the speed controller for the RSM is not capable to compensate for the large torque/current ripples induced by the faulty PMSM-converter.

*2) Experiment ($E_2$):* For this experiment, again an open-switch fault in $S_1$ (phase $a$ of the PMSM) is emulated; but this time, the fault-tolerant control system (extended field-oriented control, as proposed in Sect. III-B) with extended anti-windup, flat-top modulation and optimal $i_s^d$-injection (with $\varphi_0 = 197°$) is implemented for simulation and measurement. Figure 13(b) shows the comparative simulation and measurement results. Again, simulation and measurement results match very closely. Moreover, also the THD-values $\text{THD}_{i_s^a} = 9.4\,\%$ and $\text{THD}_{i_s^a,\text{meas}} = 10.6\,\%$ are almost identical. In conclusion, the proposed modifications are also effective in real world and the outcomes of the theoretical and simulative analysis in Sect. III-B are confirmed.

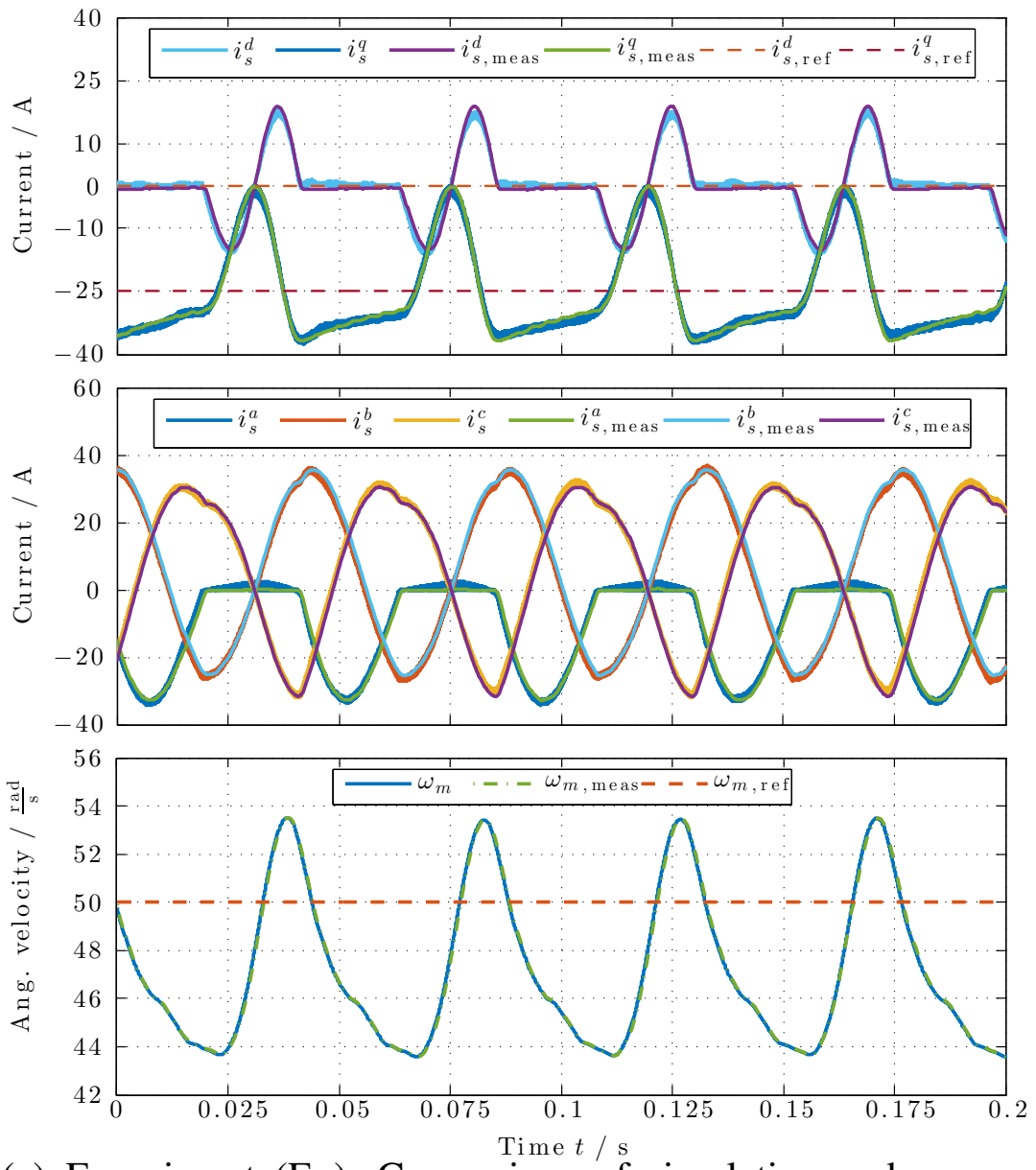

(a) Experiment ($E_1$): Comparison of simulation and measurement results for standard control system: $\mathrm{THD}_{i_s^a} = 41.8\,\%$ (simulation) vs. $\mathrm{THD}_{i_s^a,\mathrm{meas}} = 45.4\,\%$ (measurement).

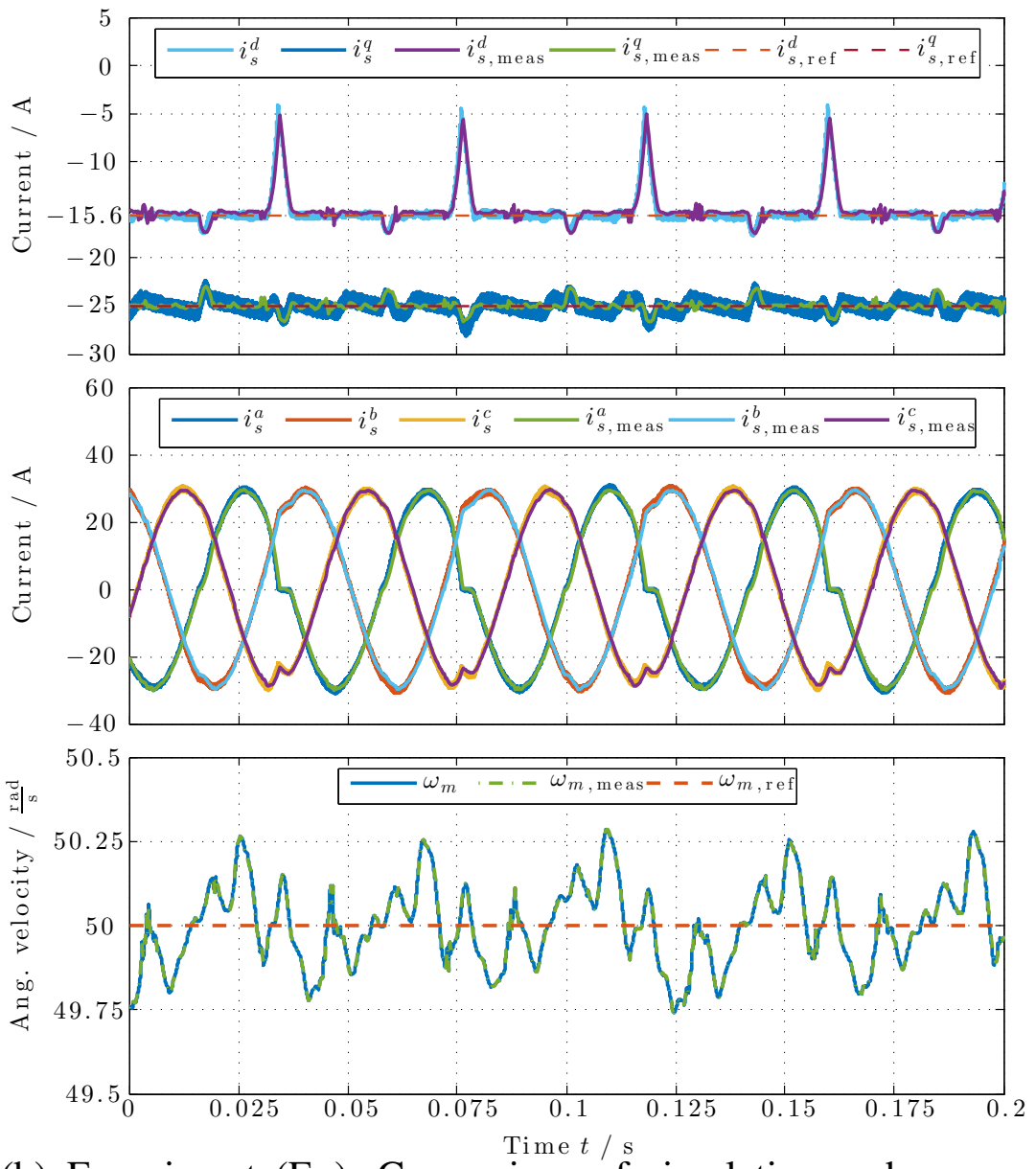

(b) Experiment ($E_2$): Comparison of simulation and measurement results for fault-tolerant control system: $\mathrm{THD}_{i_s^a} = 9.4\,\%$ (simulation) vs. $\mathrm{THD}_{i_s^a,\mathrm{meas}} = 10.6\,\%$ (measurement).

Fig. 13: *Comparison of simulation and measurement results for standard control system and fault-tolerant control system with extended anti-windup, modified SVM (flat-top modulation) and optimally injected d-current.*

*3) Experiment ($E_3$):* During the last experiment, the measurement scenario comprises a sequence of events (see Fig. 14) which are:

- 1. time interval: Standard control system (without open-switch fault/fault-free case);
- 2. time interval: Standard control system under open-switch fault in $S_1$;
- 3. time interval: Standard control system with extended anti-windup under open-switch fault in $S_1$;
- 4. time interval: Standard control system with extended anti-windup *and* modified SVM (flat-top modulation) under open-switch fault in $S_1$
- 5. time interval: Fault-tolerant control system with extended anti-windup, modified SVM (flat-top modulation) *and* optimal $d$-current injection under open-switch fault in $S_1$.

The measurement results are shown in Fig. 14. First and second subplots show the measured $d$-currents $i_\mathrm{s}^d$ & $i_\mathrm{s}^q$ with their references $i_{\mathrm{s,ref}}^d$ & $i_{\mathrm{s,ref}}^q$, and the phase current $i_\mathrm{s}^a$ with its reference $i_{\mathrm{s,ref}}^a$, respectively. The third subplot shows the rotational speed and its reference. As soon as the open-switch fault in $S_1$ occurs, the currents $i_\mathrm{s}^d$ and $i_\mathrm{s}^q$ start to oscillate within a band of 35 A. The phase current $i_\mathrm{s}^a$ cannot track its reference, since positive half-waves cannot be reproduced. The resulting torque ripples in the PMSM deteriorate the speed control of the RSM and the rotational speed begins to oscillate with an amplitude of roughly $10\,\frac{\mathrm{rad}}{\mathrm{s}}$. After the extended anti-windup strategy is enabled, the band of the current speed oscillations decreases slightly and the phase current $i_\mathrm{s}^a$ can roughly track the negative reference half-waves again. The additional use of the modified SVM (flat-top modulation) reduces the oscillation band further and more significantly and improves the current tracking capability further. Note that $i_\mathrm{s}^q$ does not approach zero anymore. Finally, enabling the optimal $i_\mathrm{s}^d$-injection achieves that $i_\mathrm{s}^d$ can again track the current reference $i_{\mathrm{s,ref}}^q$ with only very small ripples. Due to the optimal $d$-current injection, $i_\mathrm{s}^d$ is still oscillating around its reference value within a band of 13 A and the resulting magnitude of the phase current $i_\mathrm{s}^a$ is increased by approximately $17,9\%$ to 29.5 A. But the torque ripples are drastically reduced and the speed control performance is almost as good as it was in the fault-free case.

## V. Conclusion

In this paper, a generic mathematical model of a two-level converter with open-switch faults has been derived. The output voltages of the faulty converter can be computed based on switching vector, dc-link voltage and sign of the current in the phase with the open-switch fault. The model holds for faults in each of the six switches. In a next step, the impact of an open-switch fault on the current control system of isotropic permanent-magnet synchronous generators has been investigated. If the faulty switch is known and the corresponding diode is still working properly, three easy-to-implement extension to the control system have been proposed to improve fault-tolerance and control performance under faults. The three modifications reduce the THD of the faulty phase current, the torque ripples and, therefore, the stress on the mechanical drive train. Moreover, a safe and uninterrupted operation of the generator can be guaranteed. The proposed modifications are (i) extension of the anti wind-up strategy in the current PI controllers, (ii) modification of the SVM (flat-top modulation) and (iii) injection of optimal $d$-currents. All modifications have been explicitly illustrated and implemented for an open-switch fault in phase $a$; however, the generic model and the provided descriptions allow to implement the modifications for any other open-switch fault in the converter. Finally, comparative simulation and measurement results illustrate and validate (i) the accuracy of the proposed model of the faulty converter and (ii) the effectiveness and functionality of the proposed modifications on a laboratory test bench.

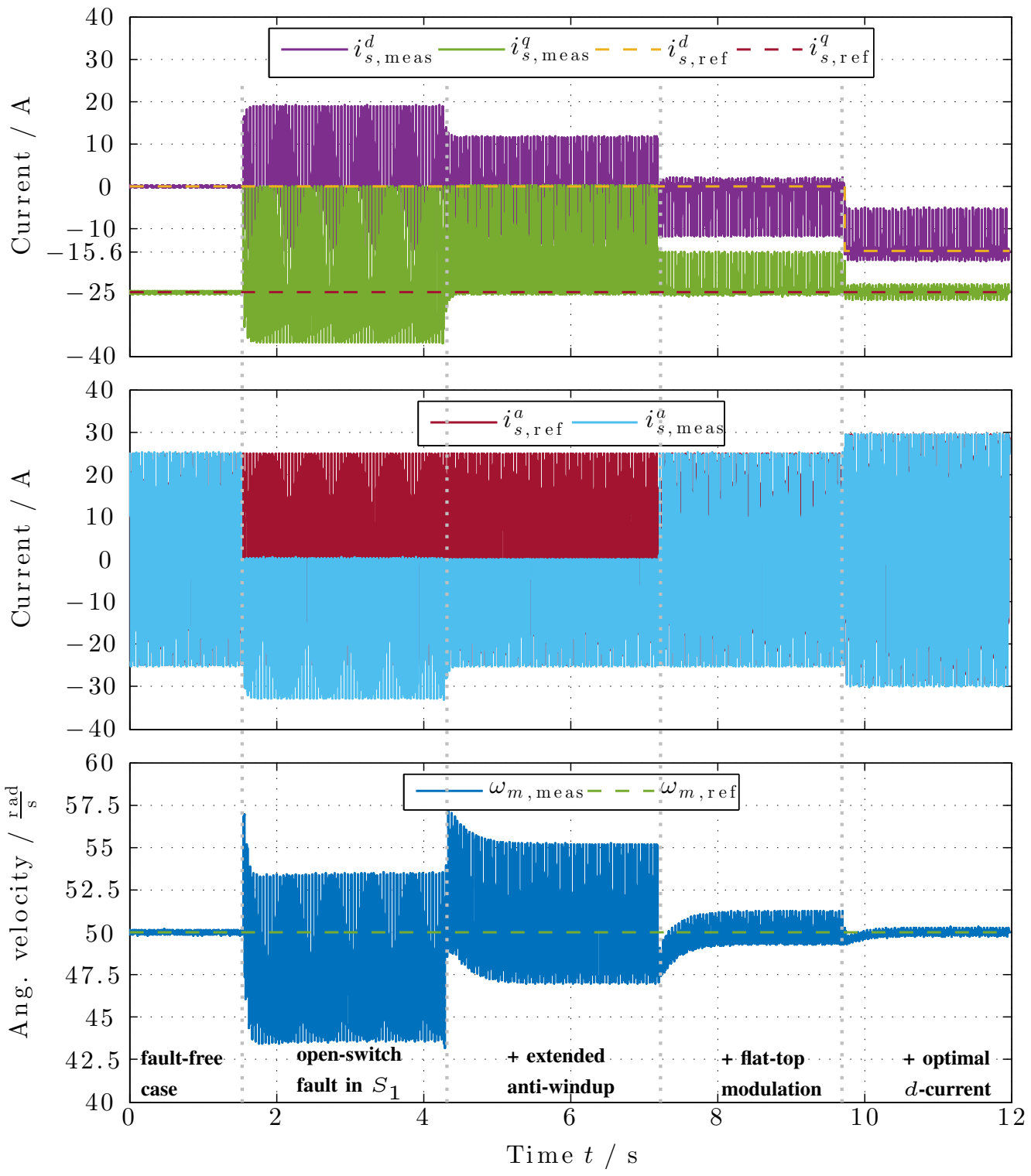

Fig. 14: *Experiment ($E_3$): Measurement results for current control system during four different scenarios: (i) fault-free case, (ii) open-switch fault in $S_1$, (iii) with extended anti-windup, (iv) additionally with flat-top modulation and (v) additionally with optimal $d$-current injection.*

## Acknowledgement

The authors are deeply indebted to Max Lindner for adapting the laboratory test bench to make measurements of a converter with open-switch faults possible. This project has received funding from the Bavarian Ministry for Education, Culture, Science, and Art.

## References

[1] B. Lu and S. K. Sharma, "A literature review of IGBT fault diagnostic and protection methods for power inverters," *IEEE Transactions on Industry Applications*, vol. 45, pp. 1770–1777, Sept 2009.

[2] K. Rothenhagen and F. W. Fuchs, "Performance of diagnosis methods for IGBT open circuit faults in three phase voltage source inverters for AC variable speed drives," in *2005 European Conference on Power Electronics and Applications*, pp. 10 pp.–P.7, Sept 2005.

[3] U. M. Choi, J. S. Lee, F. Blaabjerg, and K. B. Lee, "Open-circuit fault diagnosis and fault-tolerant control for a grid-connected NPC inverter," *IEEE Transactions on Power Electronics*, vol. 31, pp. 7234–7247, Oct 2016.

[4] J. S. Lee, K. B. Lee, and F. Blaabjerg, "Open-switch fault detection method of a back-to-back converter using NPC topology for wind turbine systems," *IEEE Transactions on Industry Applications*, vol. 51, pp. 325–335, Jan 2015.

[5] U. M. Choi, K. B. Lee, and F. Blaabjerg, "Diagnosis and tolerant strategy of an open-switch fault for T-type three-level inverter systems," *IEEE Transactions on Industry Applications*, vol. 50, pp. 495–508, Jan 2014.

[6] U. M. Choi, H. G. Jeong, K. B. Lee, and F. Blaabjerg, "Method for detecting an open-switch fault in a grid-connected NPC inverter system," *IEEE Transactions on Power Electronics*, vol. 27, pp. 2726–2739, June 2012.

[7] W.-S. Im, J.-S. Kim, J.-M. Kim, D.-C. Lee, and K.-B. Lee, "Diagnosis methods for IGBT open switch fault applied to 3-phase AC/DC PWM converter," *Journal of Power Electronics*, vol. 12, no. 1, 2012.

[8] N. M. A. Freire, J. O. Estima, and A. J. M. Cardoso, "Open-circuit fault diagnosis in PMSG drives for wind turbine applications," *IEEE Transactions on Industrial Electronics*, vol. 60, pp. 3957–3967, Sept 2013.

[9] H. T. Eickhoff, R. Seebacher, A. Muetze, and E. Strangas, "Enhanced and fast detection of open switch faults in inverters for electric drives," *IEEE Transactions on Industry Applications*, vol. PP, no. 99, pp. 1–1, 2017.

[10] R. L. de Araujo Ribeiro, C. B. Jacobina, E. R. C. da Silva, and A. M. N. Lima, "Fault detection of open-switch damage in voltage-fed PWM motor drive systems," *IEEE Transactions on Power Electronics*, vol. 18, pp. 587–593, Mar 2003.

[11] K. H. Kim, D. U. Choi, B. G. Gu, and I. S. Jung, "Fault model and performance evaluation of an inverter-fed permanent magnet synchronous motor under winding shorted turn and inverter switch open," *IET Electric Power Applications*, vol. 4, pp. 214–225, April 2010.

[12] S. M. Jung, J. S. Park, H. W. Kim, K. Y. Cho, and M. J. Youn, "An MRAS-based diagnosis of open-circuit fault in PWM voltage-source inverters for PM synchronous motor drive systems," *IEEE Transactions on Power Electronics*, vol. 28, pp. 2514–2526, May 2013.

[13] W. S. Im, J. M. Kim, D. C. Lee, and K. B. Lee, "Diagnosis and fault-tolerant control of three-phase AC-DC PWM converter systems," *IEEE Transactions on Industry Applications*, vol. 49, pp. 1539–1547, July 2013.

[14] P. Sobanski and T. Orlowska-Kowalska, "Analysis of space vector modulation technique in inverter-fed fault-tolerant induction motor drive," in *2014 16th International Power Electronics and Motion Control Conference and Exposition*, pp. 1024–1029, Sept 2014.

[15] W. S. Im, J. J. Moon, J. M. Kim, D. C. Lee, and K. B. Lee, "Fault tolerant control strategy of 3-phase AC-DC PWM converter under multiple open-switch faults conditions," in *2012 Twenty-Seventh Annual IEEE Applied Power Electronics Conference and Exposition (APEC)*, pp. 789–795, Feb 2012.

[16] S. M. Jung, K. Lee, and H. W. Kim, "Post-fault operation of open-circuit fault in three-phase PWM converter," in *2014 16th International Power Electronics and Motion Control Conference and Exposition*, pp. 311–316, Sept 2014.

[17] J. S. Lee and K. B. Lee, "An open-switch fault detection method and tolerance controls based on SVM in a grid-connected T-type rectifier with unity power factor," *IEEE Transactions on Industrial Electronics*, vol. 61, pp. 7092–7104, Dec 2014.

[18] J. S. Lee and K. B. Lee, "Open-switch fault tolerance control for a three-level NPC/T-type rectifier in wind turbine systems," *IEEE Transactions on Industrial Electronics*, vol. 62, pp. 1012–1021, Feb 2015.

[19] N. M. A. Freire, *Fault-tolerant permanent magnet synchronous generator drives for wind turbine applications*. PhD thesis, University of Coimbra, 2013.

[20] A. Gaeta, G. Scelba, and A. Consoli, "Modeling and control of three-phase PMSMs under open-phase fault," *IEEE Transactions on Industry Applications*, vol. 49, pp. 74–83, Jan 2013.

[21] O. Wallmark, L. Harnefors, and O. Carlson, "Post-fault operation of fault-tolerant inverters for PMSM drives," in *2005 European Conference on Power Electronics and Applications*, pp. 10 pp.–P.11, Sept 2005.

[22] S. Bolognani, M. Zordan, and M. Zigliotto, "Experimental fault-tolerant control of a pmsm drive," *IEEE Transactions on Industrial Electronics*, vol. 47, pp. 1134–1141, Oct 2000.

[23] C. M. Hackl, *Non-identifier based adaptive control in mechatronics: Theory and Application*. No. 466 in Lecture Notes in Control and Information Sciences, Berlin: Springer International Publishing, 2017.

[24] R. De Doncker, D. W. Pulle, and A. Veltman, *Advanced Electrical Drives*. Power Systems, Berlin: Springer-Verlag, 2011.

[25] C. Dirscherl, C. Hackl, and K. Schechner, "Modellierung und Regelung von modernen Windkraftanlagen: Eine Einführung (see https://arxiv.org/abs/1703.08661 for the english translation)," in *Elektrische Antriebe – Regelung von Antriebssystemen* (D. Schröder, ed.), ch. 24, pp. 1540–1614, Springer-Verlag, 2015.

[26] D. Schröder, *Elektrische Antriebe - Regelung von Antriebssystemen*. Berlin Heidelberg: Springer-Verlag, 4 ed., 2015.

[27] D. Shmilovitz, "On the definition of total harmonic distortion and its effect on measurement interpretation," *IEEE Transactions on Power Delivery*, vol. 20, pp. 526–528, Jan 2005.

[28] R. Valentine, *Motor Control Electronics Handbook*. McGraw-Hill Handbooks, McGraw-Hill Companies, 1998.

[29] C. Hackl and K. Schechner, "Non-ideal feedforward torque control of wind turbines: Impacts on annual energy production & gross earnings," *Journal of Physics: Conference Series*, vol. 753, no. 11, p. 112010, 2016.

[30] T. Burton, D. Sharpe, N. Jenkins, and E. Bossanyi, *Wind energy handbook*. John Wiley & Sons, 2 ed., 2011.

[31] C. M. Hackl, M. J. Kamper, J. Kullick, and J. Mitchell, "Current control of reluctance synchronous machines with online adjustment of the controller parameters," in *Proceedings of the 2016 IEEE International Symposium on Industrial Electronics (ISIE 2016)*, (Santa Clara, CA, USA), pp. 153–160, Institute of Electrical and Electronics Engineers (IEEE), jun 2016.

[32] H. Eldeeb, C. M. Hackl, L. Horlbeck, and J. Kullick, "Analytical solutions for the optimal reference currents for MTPC/MTPA, MTPV and MTPF control of anisotropic synchronous machines," in *Proceedings of the IEEE International Electric Machines & Drives Conference (IEMDC 2017)*, (Miami, FL, USA), 2017.